\documentclass[final,leqno]{siamltexmm}

\usepackage{amsmath,amssymb}
\usepackage{graphicx}% Include figure files
\usepackage{dcolumn}% Align table columns on decimal point
\usepackage{bm}% bold math
\usepackage{overpic,rotating}
\usepackage{gensymb}
\usepackage{hyperref}
\hypersetup{ 
	pdfborder = {0 0 0}
}

\usepackage{amsthm}  % for \nopunct

\usepackage{amsmath,amsfonts,amscd,amssymb}
\usepackage{cancel}

\usepackage{latexsym}
\usepackage{overpic}  
\usepackage{epstopdf}
\usepackage{rotating,color}

\graphicspath{{figures/}}

\usepackage{textcomp,xcolor}
\definecolor{MatlabCellColour}{RGB}{250,250,250}
\definecolor{MatPurp}{rgb}{.625,.1406,.9375}
\usepackage{listings}

\lstdefinestyle{customc}{
  belowcaptionskip=.25\baselineskip,
  breaklines=true,
  frame=L,
  xleftmargin=\parindent,
  language=Matlab,
  showstringspaces=false,
  basicstyle=\small\ttfamily,
  keywordstyle=\bfseries\color{white!30!black},
  identifierstyle=\color{blue},  
  commentstyle=\itshape\color{green!60!black},
  stringstyle=\color{MatPurp},
  backgroundcolor=\color{MatlabCellColour}
 }
\DeclareMathAlphabet{\mathpzc}{OT1}{pzc}{m}{it}
\newcommand\be {\begin{enumerate} }
\newcommand\ee {\end{enumerate} }

\title{Biological Mechanisms for Learning: A Computational Model of Olfactory Learning in the {\normalfont\bfseries\textit{Manduca sexta}} Moth, with Applications to Neural Nets } 

\author{Charles B. Delahunt\thanks{Department of Electrical Engineering, University of Washington, Seattle, WA, USA. delahunt@uw.edu}  \and
Jeffrey A. Riffell\thanks{Department of Biology, University of Washington, Seattle, WA, USA. jriffell@uw.edu}  \and
J. Nathan Kutz\thanks{Department of Applied Mathematics, University of Washington, Seattle, WA, USA. kutz@uw.edu} 
}

\begin{document}
\maketitle
\newcommand{\slugmaster}{%
\slugger{sifin}{}{}{}{}}%slugger should be set to juq, siads, sifin, or siims

%                                                                                                            							ABSTRACT
\begin{abstract}
The insect olfactory system, which includes the antennal lobe (AL), mushroom body (MB), and ancillary structures,  
is a relatively simple neural system capable of learning. 
Its structural features, which are widespread in biological neural systems, process olfactory stimuli 
through a cascade of networks where large dimension shifts occur from stage to stage and where sparsity
and randomness play a critical role in coding.  
Learning is partly enabled by a neuromodulatory reward mechanism of octopamine stimulation of the AL, whose increased activity induces 
rewiring of the MB through Hebbian plasticity.
Enforced sparsity in the MB focuses Hebbian growth on neurons that are the most important for the representation of the learned odor.
Based upon current biophysical knowledge, we have constructed an end-to-end computational model of the \textit{Manduca sexta}
moth olfactory system which includes the interaction of the AL and MB under octopamine stimulation.
Our model is able to robustly learn new odors, and our simulations of integrate-and-fire
neurons match the statistical features of \textit{in vivo} firing rate data.
From a biological perspective, the model provides a valuable tool for examining the role of neuromodulators, like octopamine, in learning, and gives insight into critical interactions between sparsity, Hebbian growth, and stimulation during learning.
Our simulations also inform predictions about structural details of the olfactory system that are not currently well-characterized.
From a machine learning perspective, the model yields bio-inspired mechanisms that are potentially useful in constructing  neural nets for rapid learning from very few samples.  
These mechanisms include high-noise layers, sparse layers as noise filters, and a biologically-plausible optimization method to train the network based on octopamine stimulation, sparse layers, and Hebbian growth.
\end{abstract}

%%%%%%%%%%%%%%%%%%%%%%%%%%%%%%%%%%%%%%%%%%%%%%%%%%%%%%%%%%%%%
%		INTRODUCTION               %%%%%%%%%%%%%%%%%%%%%%%%%%%%%%%%%%%%%%%%%%%
%%%%%%%%%%%%%%%%%%%%%%%%%%%%%%%%%%%%%%%%%%%%%%%%%%%%%%%%%%%%%

%\clearpage                                          

\section{Introduction}~\\
Learning is a vital function of biological neural networks, yet the underlying biomechanical mechanisms responsible for robust and rapid learning are not well understood.  
The insect olfactory network, and the moth's olfactory network (MON) in particular (e.g. the {\em Manduca sexta} moth),  is a comparatively simple biological neural network capable of learning \cite{hammerMenzel1998, Riffell2008}, and makes an ideal model organism for characterizing the mechanics of learning.
It is amenable to interrogation through experimental neural recordings of key, well-understood structural components including the antenna lobe (AL) \cite{wilson2008} and mushroom body (MB) \cite{campbellMushroomBody}.  
In addition, the AL-MB contain many structural motifs that are widespread in biological neural systems. 
 These motifs include:   
(i) the use of neuromodulators (octopamine and dopamine) in learning \cite{dacksRiffell2009}, 
(ii) a cascading networks structure \cite{masse2009},  
(iii) large changes in dimensionality (i.e. numbers of neurons) between networks \cite{laurent2002}, 
(iv) sparse encodings of data in high-dimensional networks \cite{honeggerTurner2011}, 
(v) random connections \cite{caron2016}, and 
(vi) the presence of noisy signals \cite{galizia2014}. 
Bio-inspired design principles suggest that each of the features has high value to the olfactory system.  
The mechanism of octopamine/dopamine release during learning is of particular interest, 
%both biologically and in the context of machine learning (ML), 
 since it is not well-understood how this stimulation promotes the construction of new sparse codes in the MB.   

In this work, we build a computational model of the MON that is closely aligned with both the known biophysics of the moth AL-MB and \textit{in vivo} neural firing rate data, and that includes the dynamics of octopamine stimulation.
% info re other computational models:
There exist computational neural network models of the insect MB \cite{peng2017, roper2017, mosqueiro2014, arena2013, faghihi2017} inspired by but not closely tied to particular organisms (indeed, \cite{peng2017} points out the advantages of this more general approach). 
Our goal here is to analyze the interactions of AL, MB, and octopamine during associative learning.
To this end, we model the architecture and neural dynamics of the whole system: AL, MB, octopamine stimulation, synaptic growth, odor projection onto the AL, and an extrinsic (readout) neuron downstream.
We model neurons as ``integrate-and-fire" units \cite{dayan2001}.
In addition, we tether the model closely to the known biophysics of the {\em M. sexta} moth.
This allows us to calibrate the model's behavior to \textit{in vivo} neural firing rate data from moth AL neurons, collected during odor and octopamine stimulation.   
% The simulations allow us to analyse how learning occurs in a biological neural net (NN).

We thus create a full, end-to-end neural network model (hereafter ``Network Model") that demonstrates robust learning behavior while also tightly matching the structure and behavior of a real biological system. 
This approach has three advantages:  
(i) we can meaningfully compare Network Model simulation output to experimental data in order to tune model parameters, 
(ii) simulation results can map back to the original biological system in order to render meaningful insights, and 
(iii) each structural and dynamical feature of the system can demonstrate its relevance to the task of learning in a neural net. 
\textit{In silico} experiments with the Network Model allow us to abstract out critical features in the moth's toolkit, study how these features enable learning in a neural network, and thus derive bio-inspired insight into the mathematical framework that enables rapid and robust learning.    
%This model overall provides a new paradigm for engineering neural nets (NN) with the goal of robust and rapid learning.
% Our experiments demonstrate how the components of the MON (schematic in Fig. \ref{schematicPlusTimecourses} A) produce robust and rapid learning, thus providing key biological insights into learning and offering bio-inspired design principles for neural networks more broadly.
Specifically, our experiments elucidate mechanisms for fast learning from noisy data that rely on cascaded networks, sparsity, and Hebbian plasticity. 

These mechanisms have potential applications to engineered neural nets.
NNs have emerged as a dominant mathematical paradigm for characterizing neural processing and learning.  
This is not surprising given that they were inspired by the Nobel prize winning work of Hubel and Wiesel on the primary visual cortex of cats \cite{hubel}.  
These seminal experiments showed that networks of neurons were organized in hierarchical layers of cells for processing visual stimulus.  
The first mathematical model of a neural network, the {\em Neocognitron} in 1980 \cite{fukushima}, had many of the characteristic features of today's  deep neural networks  (DNNs).   
However, many of the biological motifs listed above are largely absent from engineered NNs, whose principles and building blocks are biologically implausible even as DNNs have achieved great success \cite{schmidhuber2014, lecunIeeeSpectrum}. 
%Although variations of neural networks have been theoretically posited to characterize learning,
For example, the AL-MB interaction with octopamine and Hebbian plasticity operates in a fundamentally different manner than the backpropagation optimization used in DNNs.
We seek to characterize a set of biological elements, a ``biological toolkit", that can be assembled into NNs that operate on fundamentally different principles than standard NNs and that are capable of rapid and robust learning from very few training samples, an ability common in biological NNs that remains challenging for today's DNNs.  

\begin{figure}[h!]
%\begin{adjustwidth} {-55mm}{0mm}
\centering
\includegraphics [width=120mm] {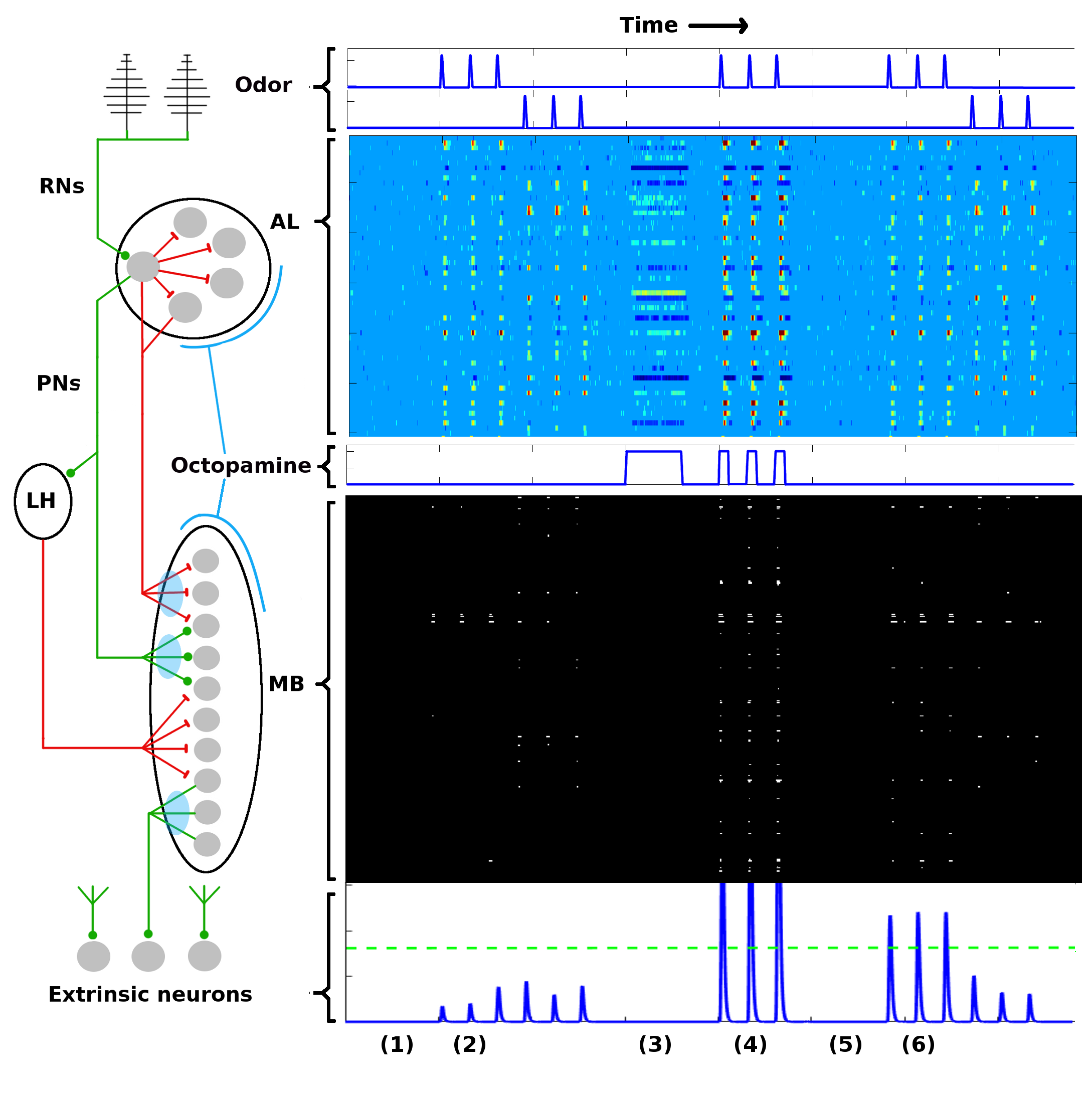}
% \begin{overpic} [width=190mm] {figsForPlosPaper_nov/pngPlaceholders/overviewFig_v3.png}  
%\put(55,95){Time $t$ $\longrightarrow$}
%\end{overpic}
%
\vspace*{-.2in}
\caption{ {\bf AL-MB overview.}
{\bf{A:}}
System schematic: 
Chemical sensors (RNs) excite a noisy pre-amp network (AL), which  feeds forward to a plastic sparse memory layer (MB), which excites readout (decision) neurons (ENs). Green lines show excitatory connections, red lines show inhibitory connections (LH inhibition of the MB is global). 
Light blue ovals show plastic synaptic connections into and out of the MB.  \newline
{\bf{B:}}
Neuron timecourse outputs from each network (typical simulation) with time axes aligned vertically. 
Timecourses are aligned horizontally with their regions-of-origin in the schematic. 
The AL timecourse shows all responses within $\pm$ 2.5 std dev of mean spontaneous rate as medium blue. 
Responses outside this envelope are yellow-red (excited) or dark blue (inhibited). 
MB responses are shown as binary (active/silent). 
Timecourse events are as follows: 
(1) A period of no stimulus. All regions are silent.
(2) Two odor stimuli are delivered, 3 stimulations each. AL, MB, and ENs display odor-specific responses. 
(3) A period of control octopamine, ie without odor or Hebbian training. 
AL response is varied, MB and EN are silent.
(4) The system is  trained (octopamine injected) on the first odor. 
All regions respond strongly.
(5) A period of no stimulus. 
All regions are silent, as in (1).
(6) The stimuli are re-applied. 
The AL returns to its pre-trained activity since it is not plastic. 
In contrast, the MB and EN are now more responsive to the trained odor, while response to the untrained odor is unchanged. 
Green dotted line in the EN represents a hypothetical ``action" threshold. 
The moth has learned to respond to the trained odor. 
  }
\label{schematicPlusTimecourses}
%\end{adjustwidth}
\end{figure}

A brief summary of the AL-MB network:
It is organized as a feed-forward cascade of five distinct networks, as well as a reward mechanism  \cite{martin2011, kvello2009}.
Roughly 30,000 noisy chemical receptor neurons (RNs) detect odor and send signals to the Antenna Lobe (AL) \cite{masse2009}.
The AL acts as a pre-amplifier, providing gain control and sharpening of odor representations \cite{bhandawat2007}.
It contains roughly 60 isolated units (glomeruli), each focused on a single odor stimuli feature  \cite{martin2011}.
Glomeruli laterally inhibit each other, and project odor codes to the Mushroom Body (MB).
 AL neurons are noisy  \cite{galizia2014}.
The MB contains about 4000 Kenyon Cells (KCs).
These fire sparsely and  encode odor signatures as memories \cite{perisse2013, campbellMushroomBody, honeggerTurner2011}.
MB sparsity is enforced by global inhibition from the Lateral Horn \cite{bazhenovStopfer2010}.
Extrinsic Neurons (ENs), numbering $\sim$10's, are believed to be ``readout neurons" that interpret the KC codes \cite{campbell2013, hige2015}.
In response to reward (sugar at the proboscis), a large neuron sprays octopamine globally over the AL and MB, causing generalized stimulation of neurons.
Learning does not occur without this octopamine input  \cite{hammer1995, hammerMenzel1998}.
The connections into the KCs (AL$\rightarrow$KCs) and out of the KCs (KCs$\rightarrow$ENs) are  plastic during learning  \cite{cassenaer, masse2009}. 
Figure \ref{schematicPlusTimecourses} gives a system schematic (A) along with typical firing rate (FR) timecourses (from simulation) for neurons in each network (B).
More network details are given in Methods. \newline

%----------------------------------------------------------------------------------------------------------------------------------------------------------------------------------------------
%----------------------------------------------------------------------------------------------------------------------------------------------------------------------------------------------

\section{Results}~\\
We first show the calibration of our Network Model to \textit{in vivo} data.
We then describe neural behaviors of the Network Model during learning and give results of learning experiments.
Finally, we give results of experiments on MB sparsity.

\subsection{Calibration of Model}
\begin{figure}[t]
%\begin{adjustwidth} {-50mm}{0mm}
\centering
\includegraphics [width=120mm] {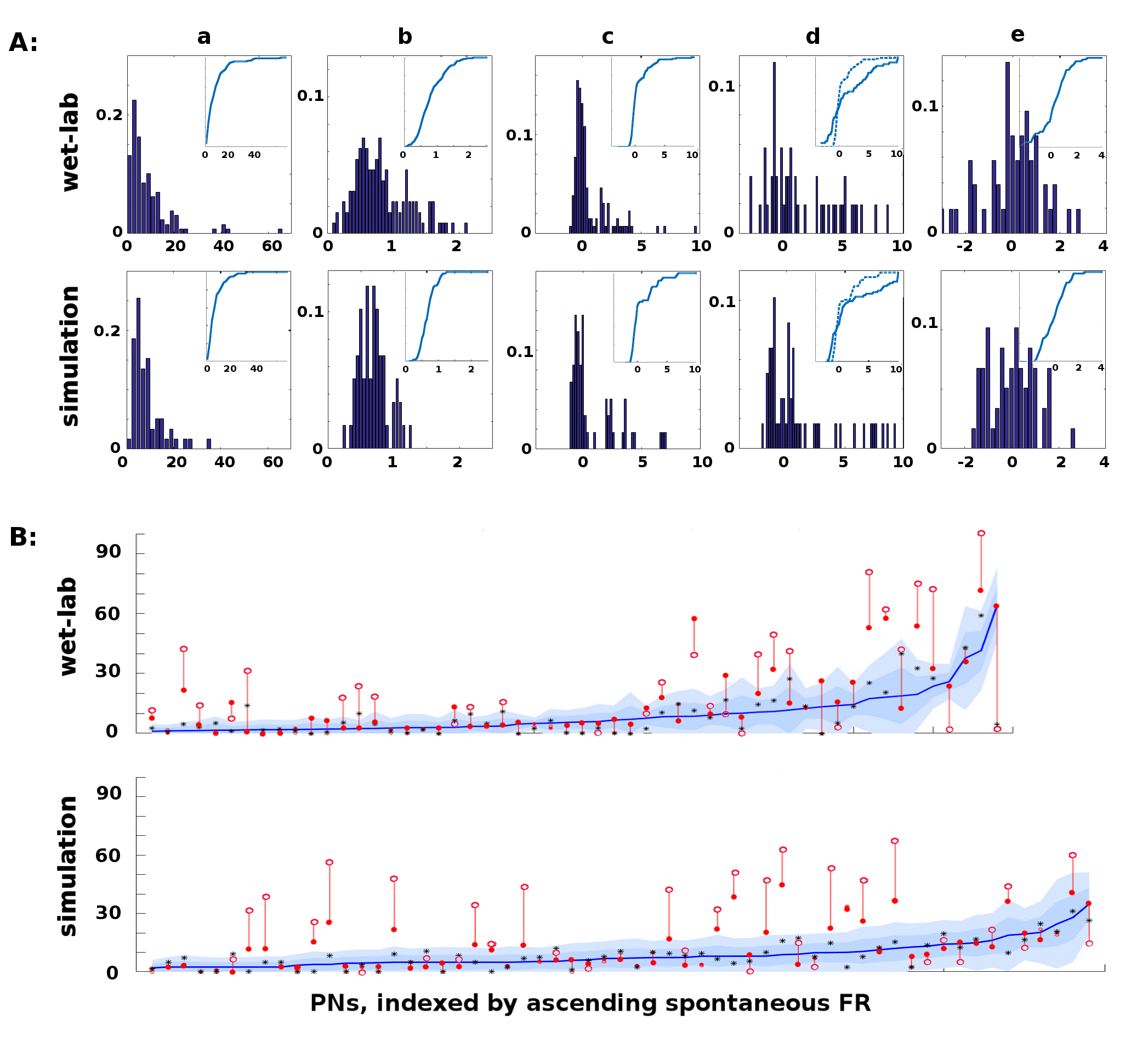}   
\caption{ {\bf In vivo firing rate data and model calibration:}  Comparison of PN firing rate activity from \textit{in vivo} data and simulations. 
{\bf A:} Histograms and CDFs of \textit{in vivo} data and simulations.  
{\bf Col a}: mean spontaneous FRs $\mu_s$. 
{\bf Col b}: $\sigma_s / \mu_s$ of spontaneous FRs, a measure of noisiness of a PN.  
{\bf Col c}: odor response, measured as distance from $\mu_s$ in $\sigma_s$ units.  Distance $>2\sigma_s$ implies a strong activation/inhibition.  
{\bf Col d}: odor response during octopamine,  in $\sigma_s$ units distance from $\mu_s$. 
Note that PN responses are broadened (i.e. more PNs are strongly activated or inhibited).
The dotted line in the CDF inset is the same as the CDF of the odor response without octopamine, to show the broadening towards both extremes. 
{\bf Col e}: change in mean spontaneous FRs due to octopamine, measured in $\sigma_s$ units distance from (non-octopamine) $\mu_s$. 
Some PNs are excited, some are inhibited.\newline 
{\bf B:} Activity of PNs indexed by increasing spontaneous FR.  
Blue lines = mean spontaneous FRs $\mu_s$ (cf col 1). Shaded regions = $\sigma_s, 2\sigma_s$ envelopes (cf col 2). 
Solid red dots = odor response FRs (cf col 3). 
Hollow red dots = odor response FRs during octopamine (cf col 4). 
Red lines show the change in odor response FRs due to octopamine (cf broadened response). 
Black stars (*) = spontaneous FRs during octopamine (cf col 5). 
In panel A cols 3, 4, 5, the x-axes are expressed in units of $\sigma_s$, while in panel B the y-axis measures raw spikes/sec FR.   
}
\label{comparisonPlots}
%\end{adjustwidth}
\end{figure}

~\\
The Network Model was calibrated to behave in a statistically similar way to three  sets of \textit{in vivo} data measuring  projection neuron (PN) firing rate (FR) activity in the AL (See Methods for details):  
(i) PN spike counts with odor but without octopamine: 129 units with FR$>$1 spike/sec, (ii)  PN spike counts with odor and with octopamine: 180 units with FR$>$1 spike/sec, and (iii) PN spike counts with odor, with and without  octopamine: 52 units with FR$>$1 spike/sec.  
Due to the limited number of experimental units, only qualitative comparisons of the model and experiment could be made.  
Excessive tuning of the model parameters  would have served only to overfit the particular data, rather than matching true PN behavior distributions or, more importantly, the general learning behavior of the moth.   
Fig \ref{comparisonPlots} shows the close match of typical Network Model PN statistics to \textit{in vivo} PN statistics based on mean ($\mu$) and variance ($\sigma$) of spontaneous FRs and odor responses, both without and with octopamine  
(details of metrics are given in Methods).
Importantly, Fig \ref{comparisonPlots} shows significant octopamine-modulated increase in PN FR activity in the Network Model, consistent with \textit{in vivo} experiments involving octopamine stimulation.

There is limited experimental data measuring the FR activity of Kenyon cells (KC) in the MB, and no data to our knowledge measuring KC in response to octopamine stimulation.  
However, we note that the behavior of KCs during the application of octopamine to the AL, either with or without odor, is not an artifact of parameter tuning.  
Rather, it follows from the tuning the AL to match \textit{in vivo} data.  
Specifically, PN FRs at baseline (with no odor or octopamine), with odor alone, with octopamine alone, and with odor and octopamine, are all determined by calibration of the model to \textit{in vivo} data.  
KCs respond only to PNs and to inhibition from the LH (See Fig \ref{schematicPlusTimecourses}).  
Calibrating the KC baseline response in the absence of octopamine to \textit{in vivo} data in \cite{turner2008} fixes the feed-forward connections from PNs. 
% The KC behavior is then determined entirely by the PN dynamics.  
Assumed in this model, due to lack of biophysical evidence, is that octopamine has no direct effect on KC FRs.  
Thus KC behavior with octopamine is fully determined once the model is tuned to PN data.  
This completes the calibration process of our model parameters.  
As Fig \ref{comparisonPlots} shows, the model agrees well with \textit{in vivo} experiment.

There are no bulk data, to our knowledge, measuring EN firing rates in response to odors and/or octopamine.
However, calibrating EN response is not necessary to demonstrate an ability to learn. 
The key marker is post-training increase in EN response.

%\clearpage

\subsection{Learning Experiments: PN and KC Behaviors}~\\
PN activity in the AL, and KC activity in the MB, from typical Network Model simulations  are shown in Fig \ref{schematicPlusTimecourses} as heatmaps, evolved over the time course of a simulation in which the system was exposed to two different odors and trained on one of them. 
The AL is stimulated with octopamine during training.
Each row of the heatmap represents a distinct PN or KC as it evolves in time (left to right columns of heat map). 
All the timescales are aligned. 
 
%PN activity in the AL is shown in  Fig \ref{schematicPlusTimecourses} as a heatmap.  
%In the simulations, the output of 60 PNs are monitored over time as the AL is stimulated with two different odors and trained on one of them. 
\paragraph*{PNs} In the AL heatmap, the light blue region corresponds to PN FRs within 2.5$\sigma_s$ of their respective mean spontaneous FRs $\mu_s$, warm colors correspond to very high FRs, and dark blues correspond to strongly inhibited FRs.  
The simulations demonstrate a number of key PN behaviors, including  
(i) absent odor, PN FRs stay within their noise envelopes (by definition), 
(ii) the two odors have distinct excitation/inhibition signatures on PNs, 
(iii) octopamine alone (without odor) results in more PNs being excited beyond their usual noise envelopes, and also results in some PNs being inhibited beyond their usual envelopes, 
(iv) octopamine and odor, applied together, result in an overall excitation of PNs, and 
(v) the AL behavior returns to baseline after octopamine is withdrawn, since AL connection weights are not plastic.  
This last observation is in contrast to what occurs in the MB.

\paragraph*{KCs}In the MB, the KCs fire sparsely due to global inhibition from the Lateral Horn.  
The only plastic connections in the AL-MB system involve the KCs:  
Between PNs (and QNs) and KCs ($M^{PK}, M^{QK}$ in Methods Section); and between KCs and extrinsic readout neurons (ENs) ($M^{KE}$ in Methods Section).  
Thus the KC odor signatures are modulated with training.  
% As for the PNs, Fig \ref{schematicPlusTimecourses} shows a heat map of the KC FR activity over the time course of the simulation.  The simulation is again for two odors, one of which is reinforced with octopamine. 
% Each row is one of the 500 randomly selected KCs in the simulation as it evolves in time (left to right columns of heat map).  
Black regions indicate FRs $<$ 1 spike/sec, white regions  indicate FRs $>$ 1 spike/sec. 
The white regions have been dilated to make the sparsely-firing KCs easier to see.  

The simulations demonstrate a number of key KC behaviors, including 
(i) the baseline KC FR response absent any odor is essentially zero, 
(ii) the two odors excite distinct sets of KCs with varying consistency from noise trial to noise trial, 
(iii) for a given odor, some KCs fire reliably in response to an odor stimulation and some fire only occasionally, 
(iv) when subject to octopamine but no odor, KCs are unresponsive, a benefit during learning since it prevents environmental noise from being encoded as meaningful, 
(v) when subject to octopamine and odor, KCs respond strongly to the odor with high trial-to-trial consistency, and 
(vi) the global inhibition from the LH controls the level of sparseness in the KCs, both their silence absent any odor (with or without octopamine), and their sparse firing in response to odors.  

\begin{figure}[t]
%\begin{adjustwidth} {-60mm}{0mm}
\centering
\includegraphics [width=120mm] {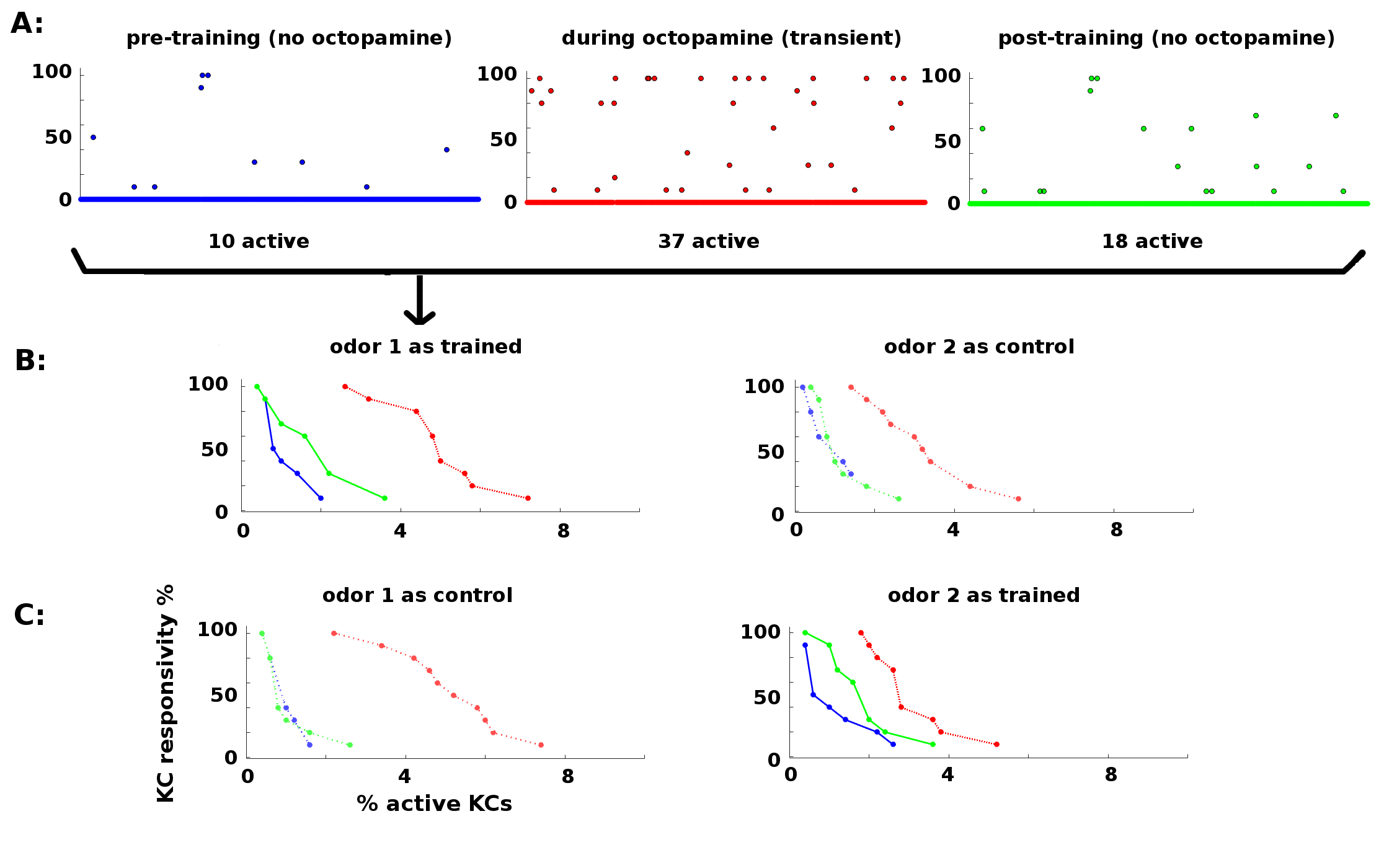}   
\caption{ {\bf KC responses to odor during training:}
KCs respond sparsely to odor pre- and post-training, ie absent octopamine (blue and green dots and curves). 
Octopamine induces transient increased responsivity (red dots and curves). 
Training results in permanent increases in response to the trained odor, but no increase in response to control odor (green dots and curves).\newline
{\bf{A:}}
KC response to an odor before, during, and after training. 
x-axis: indexed KCs (500 shown). 
y-axis: consistency of response (in \%). 
The plots are for odor 1 as the trained odor (ie same data as panel B). 
Blue = pre-training (no octopamine). 
Red = during training (with octopamine); note the heightened transient response. 
Green = post-training (no octopamine). 
There is a permanent increase in the number of KCs that respond to the trained odor.\newline
{\bf{B:}}
Response rate vs. percentage of active KCs for trained and control odors before, during, and after training. \newline
x-axis: percentage of KCs responding at the given rate. 
y-axis: consistency of response (in \%). 
Blue = pre-training. 
Red = during octopamine (transient). 
Green = post-training. 
The LH plot shows odor 1 as the reinforced odor. 
The scatterplots in (A) correspond to the three curves in this plot. 
Note that the permanent KC response curve shifts up and to the right (blue$\rightarrow$green) in the trained odor, ie more KCs respond to the odor (right shift) and they respond more consistently (upward shift). 
The RH plot shows odor 2 as a control. The control's permanent KC response curve does not shift.\newline
{\bf{C}}: 
As (B) above, but in this experiment odor 1 is now the control (LH plot), and odor 2 is reinforced  (RH plot). 
In this case, the response curve of odor 2 (reinforced) shifts to the right (blue$\rightarrow$green), while the response curve of odor 1 (control) is unchanged.  }
\label{kcConsistency}
%\end{adjustwidth}
\end{figure}

Statistics of KC responses to odors pre-, during, and post-training are shown in Fig \ref{kcConsistency}.
Naive moths have low KC response to odors, in both percentage of KCs activated and their consistency of response to odor stimulations (Fig \ref{kcConsistency},  blue dots and curves).
During training octopamine induces high KC response,  in both percentage and consistency (Fig \ref{kcConsistency}, red dots and curves).
After octopamine is withdrawn, KC response is lower than during training, but remains higher than naive levels in both percentage and consistency (Fig \ref{kcConsistency},  green dots and curves) for the trained odor only.
Thus the newly-learned importance of the trained odor is encoded as broader and stronger KC responses by means of strengthened synaptic connections.%\newline

EN (readout neuron) activity is also shown over time at the bottom of Fig \ref{schematicPlusTimecourses}.
Learning is evidenced by the increased EN response to the trained odor even after octopamine has been withdrawn, due to Hebbian growth of synaptic connections into and out of the MB.

The FR activity of the PNs in the AL,  the KCs in the MB, and the ENs, as illustrated in Figs. \ref{schematicPlusTimecourses} and \ref{kcConsistency}, demonstrate the entire learning process that occurs under the influence of octopamine stimulation.  
Without octopamine learning  does not occur.  
Interestingly, although the AL does not itself experience plasticity changes, it is the AL's increased FR activity (induced by octopamine) which enables permanent rewiring of the MB via Hebbian plastic updates.   
% This is explored further in what follows.

%\clearpage
 
%--------------------------------------------------------------------------------------------------------------------------------------------------------------------------------------------
%--------------------------------------------------------------------------------------------------------------------------------------------------------------------------------------------

\subsection{Learning Experiments: EN Behavior}~\\
A key finding of this paper is that the AL-MB model demonstrates robust learning behavior.  
Here ``learning" is defined as rewiring the system so that the reinforced odor yields a significantly stronger response in the readout neuron (EN) post-training, relative to naive (i.e. pre-training) response to that odor, and also relative to the post-training responses to control odors.

\paragraph*{Learning Experiment Structure}
Moths were randomly generated from a fixed parameter template, which included randomly-assigned input maps (odor$\rightarrow$AL) for four odors.  
The odors projected broadly onto the AL, each odor hitting $\sim$20 out of 60 glomeruli.  
As a result, their projections onto the AL overlapped substantially.  
On average, a given odor uniquely projected onto about 6 glomeruli, and shared the other 14 glomeruli with other odors. 
Each generated moth was put through a series of training experiments, with each run in the series structured as follows:
\begin{enumerate}
\item The moth first received a series of stimulations from each odor, to establish a baseline (naive) EN response.
The stimulations were 0.2 seconds long and separated by gaps of several seconds.
\item The moth was trained  on one of the odors for 1 to 4 sessions (one session = 5 odor stimulations), by applying odor and octopamine concurrently.  
The MB plastic weights were updated according to a Hebbian rule. 
\item Upon completion of training,  the four odors were each again applied as a series of odor stimulations, to establish post-training  EN response.
\end{enumerate}

For each \{odor, \#sessions\} pair, this experiment was conducted 11 times (i.e. 11 noise realizations), for a total of 176 experiments on each moth.  These results were aggregated to assess the particular moth's learning response.\\

\begin{figure}[t]
%\begin{adjustwidth} {-45mm}{0mm}
\centering
 \includegraphics [width=120mm] {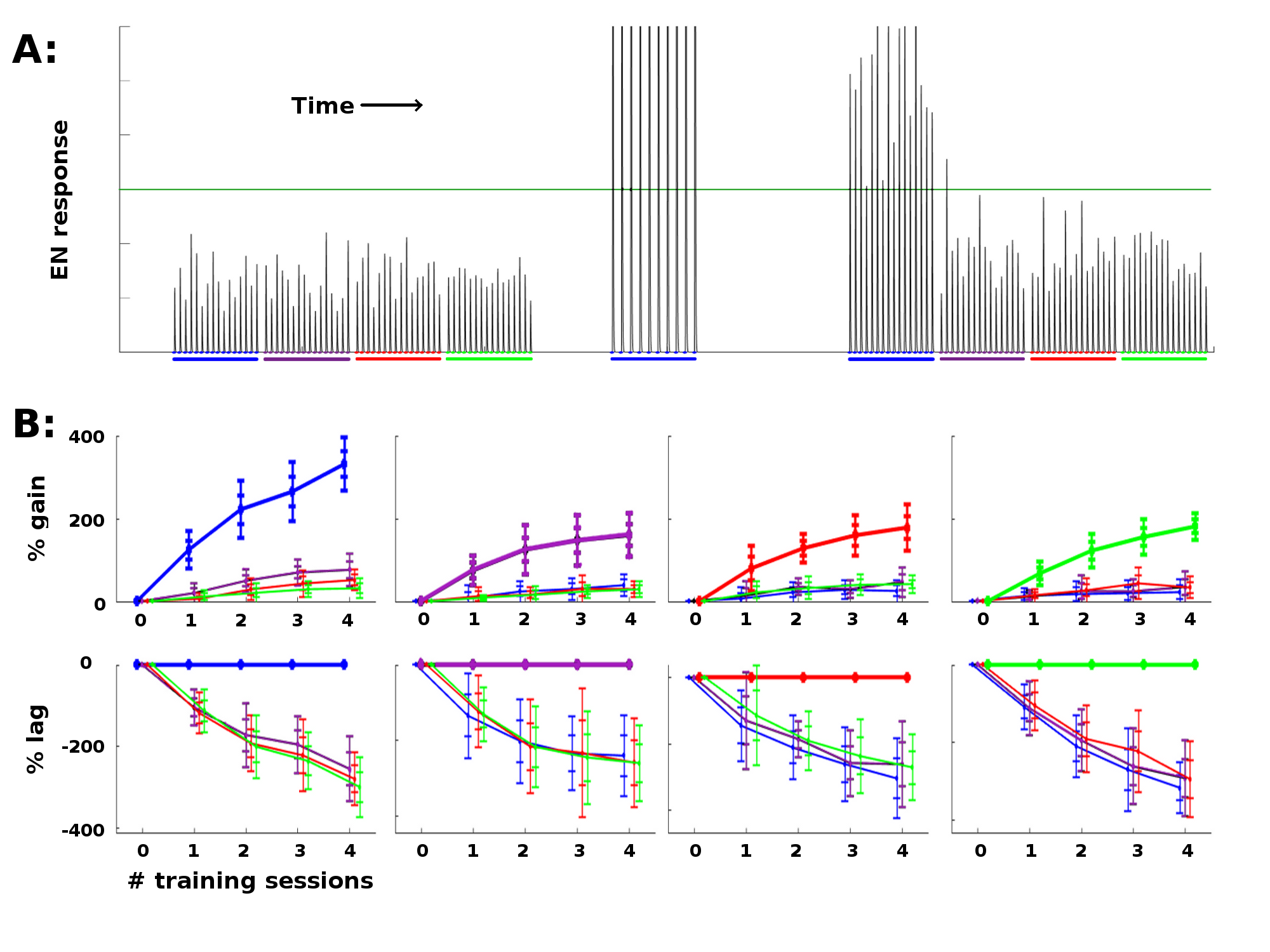}   
\caption{ {\bf Effect of training on EN FRs:}
{\bf{A:}} Typical timecourse of EN responses from an experiment with a single moth. First, 16 stimulations of each odor were delivered, to establish naive odor responses. 
Note EN response variability due to noise in the system, especially in the AL.
Next, the moth was trained on the first (blue) odor trained over 2 sessions (10 stimulations), by delivering odor and octopamine concurrently 
This timecourse corresponds to the \{odor, \#sessions\} pair in the first column in panel B, at index 2 on the x-axis.
Octopamine was then withdrawn, and the four odors were again delivered in series of stimulations, to establish post-training changes in EN response. 
The long green line represents a hypothetical trigger threshold, such that EN response $>$ threshold would induce a distinct behavior.\newline
{\bf{B:}} EN response changes due to training, aggregated results with 11 noise realizations for each \{odor, \#sessions\} pair. 
Each column shows results of training a given odor, color coded: blue, purple, red, green. 
x-axis = number of training sessions. \newline
First row: The y-axis measures percent change in EN FR.
The line shows mean percent change.
 The error bars show $\pm $1, 2 stds.\newline
Second row: The  y-axis measures percent changes in EN response, relative to the trained odor (ie subtracting the trained odor's change from all odors). 
This shows how far each control odor lags behind the trained odor. 
The line shows mean percent lag.
The error bars show $\pm $1, 2 stds.   }
\label{moth9Results}
%\end{adjustwidth}
\end{figure}

\paragraph*{Learning Experiment Results}
As a general rule, Network Model moths consistently demonstrated strong learning behavior: 
Training increased EN response to the trained odor well beyond naive levels, and also much more than it affected EN response to control odors.
 Fig \ref{moth9Results} summarizes the changes in EN responses in a typical experiment on a moth with four odors.  
Panel A shows a typical noise realization timecourse, where one odor was reinforced with octopamine and the other three odors were controls.  
Panel B shows the statistics of EN response modulation, according to \{odor, \#sessions\} pairs.

For ease of interpretation, the moth shown in Fig \ref{moth9Results} had naive EN responses of roughly equal magnitude for all four odors. 
When naive EN response magnitudes were highly uneven ($>$ 3x), robust learning still occurred, but the interpretation of the results is more complex due to scaling issues. 
A typical experiment using a moth with odor responses of highly unequal magnitude is shown in Fig \ref{unequalMag4OdorResults}. \\% LearningFig}.

\paragraph*{Points of interest (EN responses to learning)}
\begin{enumerate}
\item Because  EN response is driven solely by feed-forward signals from KCs, ENs had response $\approx$ 0 in the absence of odor, with or without octopamine, as expected (since KCs are silent absent any odor).  Thus Hebbian growth during training did not  increase EN baseline (no-odor) response.
\item The EN response to odor + octopamine was always very strong, as seen in Fig \ref{moth9Results} (A), where EN responses to odor + octopamine extend above the top of the figure.  Note that this effect follows automatically from the calibration of the Network Model to \textit{in vivo} data. Its functional value to the moth is addressed in the Discussion.
\item Training consistently increased the EN response to the reinforced odor much more than EN response to control odors, measured as percentage of naive odor response.

Since the Network Model did not include a Hebbian decay dynamic (for simplicity, absent clear evidence), this was the key indicator of robust learning.  That is, focused learning was expressed by substantially higher increase in EN response to reinforced vs control odors.  We assume that an added Hebbian decay term would have knocked smaller increases back, thus returning control odor responses to baseline.
\end{enumerate}

Results of ANOVA analysis for differential effects of training on reinforced vs unreinforced odors shows that when naive odor EN response magnitudes were similar (within 3x of each other) p-values were consistently $<$ 0.01.  
ANOVA analysis results are given in Section \ref{anova}.
% This further confirms the analysis of Figure \ref{anovaScatterplot}.

%----------------------------------------------------------------------------------------------------------------------------------------------------------------

%\clearpage

%------------------------------------------------------------------------------------------------------------------------------------------------------------------------------------------
%------------------------------------------------------------------------------------------------------------------------------------------------------------------------------------------

\subsection{MB Sparsity Experiments}~\\ \label{mbSparsity}
Projection into a high-dimensional, sparse layer is a common motif in biological neural systems \cite{ganguli2012, litwinKumarHarris2017}.
To explore the role of MB sparsity during learning, we ran Network Model experiments that varied the level of generalized inhibition imposed on the MB (the lateral horn, LH, controls MB sparsity level). 
Each experiment set a certain level of LH inhibition, then ran simulations (see Methods) that trained moths on one odor with 15 odor stimulations and left one control odor untrained.
EN responses to both trained and control odors were recorded, as well as the percentage of KCs active in response to odor.

Too little damping from the LH resulted in a high percentage of KCs being active (low sparsity).
This regime  gave consistent EN responses to odor. 
But it also caused EN responses to both control odor and noise to  increase significantly during training, reducing the contrast between EN responses to trained and control odors and also increasing spontaneous EN noise.

Too much damping resulted in a very low percentage of KCs being active (high sparsity). 
This ensured that training gains were focused on the trained odor while EN response to control odors and noise were not boosted.
However, in this regime EN responses to all odors, both pre- and post-training, were generally unreliable because too few KCs were activated.

\begin{figure}[t]
%\begin{adjustwidth} {-45mm}{0mm}
\centering
 \includegraphics [width=120mm] {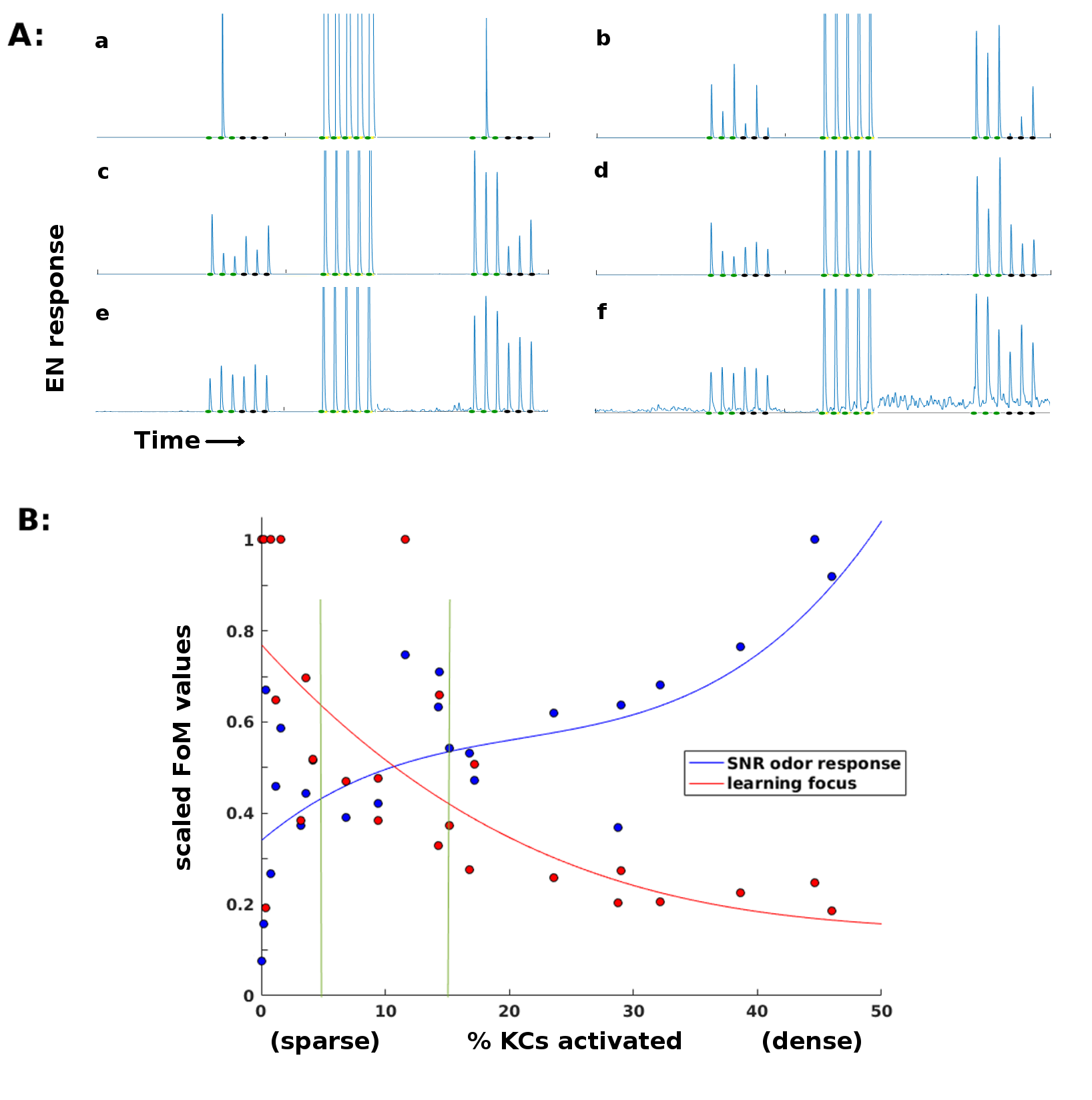}  
 \vspace*{-.2in}
\caption{ {\bf Effects of sparsity on learning and EN reliability}
 Results for a typical experiment on a moth with two odors. 
{\bf{A:}} EN responses timecourses for two odors, at varying levels of KC activation (a, b: $<$1\%. c, d: 5 to 15\%. e, f: 20 to 45\%.      
Order of events: 3 stimulations of each odor as baseline, train on first odor (only one session shown), then 3 stimulations each post-training. At very sparse levels (a, b) training is focused but odor response is not reliable. At low sparsity levels (e, f) training is unfocused, boosting EN response to control odor and to background noise.
{\bf{B:}} 
Two Figures of Merit (FoMs) plotted against MB sparsity.
Low KC activation (high sparsity) correlates with well-focused learning, but low odor response SNR. 
High KC activation (low sparsity) correlates with poorly-focused learning, but high odor response SNR.
The FoMs are each normalized for easier plotting.
y-axis: 
Blue data: $\frac{\mu(f)} {\sigma(f)}$, a measure of odor EN response SNR, where $f$ = EN odor response. 
Red data:  $ \frac{\mu(f_T)}{\mu(f_C)} $, a measure of learning focus, where $\mu(f_T)$ = mean EN post-training response to  reinforced odor; $\mu(f_C)$ = mean EN post-training response to control odor (values are thresholded at 1 for plotting).
A high value indicates that increases in EN response due to training were focused on the trained odor; low values indicate that irrelevant signal ($F_C$) was also boosted by training.
The points are experimental data, the curves are cubic fits.
Vertical green lines indicate the 5 - 15\% sparsity region, typical in biological neural systems.
}
\label{sparsityEffects}
%\end{adjustwidth}
\end{figure}

Thus sparseness in the high-dimensional MB fulfilled a vital role in the Network Model's learning system.
LH inhibition of the MB had an optimal regime that balanced opposing demands, viz for reliable odor response and for well-targeted Hebbian growth, such that on one hand EN odor responses were reliable, and on the other hand training gains were focused on the trained odor only.
Timecourses illustrating this effect are seen in Fig \ref{sparsityEffects} (A).
Fig \ref{sparsityEffects} (B) shows how this trade-off varied with MB sparsity, by plotting two figures-of-merit:
\begin{equation} \label{snrEqn}
\text{Signal-to-Noise Ratio (SNR) } = \frac{\mu(f)} {\sigma(f)} \text{ where } f  = \text{ EN odor response;~~~~~~~~~~~~~}
\end{equation}
  and 
\begin{equation}  \label{learningFocusEqn}
\text{Learning Focus }=  \frac{\mu(f_T)}{\mu(f_C)} \text{, where $\mu(f_T)$ = mean post-training EN response to }
	\end{equation} 
~~~~~~~~~~~~ trained odor, $\mu(f_C)$ = mean post-training EN response to control odor. \newline \newline

%------------------------------------------------------------------------------------------------------------------------------------------------------------------------------------------

%\clearpage
 
%%%%%%%%%%%%%%%%%%%%%%%%%%%%%%%%%%%%%%%%%%%%%

\section{Discussion}~\\
Our discussion centers around observations and implications of our Network Model experiments, which offer insights into the moth olfactory network and how it learns, including:
(i) predictions about aspects of the AL-MB still unclear in the literature, (ii) the role of sparse layers, (iii) the role of octopamine, (iv)  the value of randomness, and (v) the value of noise. 

In addition, we consider these insights in the context of machine learning algorithms.
 
%------------------------------------------------------------------------------------------------------------------------------------------------------------------------

\subsection{Predictions re details of AL-MB structure}~\\  \label{discussionPredictions}  
Model simulations enable us to make predictions about some unresolved aspects of the moth's AL-MB system.
Some examples:\newline

\paragraph*{Do LNs inhibit PNs and LNs as well as RNs} 
In the AL, LNs have a net inhibitory effect on PNs \cite{olsenWilson2010, Lei2002}, but the exact means to this end are not clear.
In particular, while LNs are known to inhibit RNs \cite{olsenWilson2010}, it is less clear whether or to what degree LNs also directly inhibit PNs and LNs.
Efforts to calibrate our Network Model to \textit{in vivo} data indicate that LNs need to inhibit not just RNs, but also (to a lesser degree) LNs and PNs.
The model weight strengths for LN$\rightarrow$RN, $\rightarrow$LN, and $\rightarrow$PN are in the ratio of 6:2:1.
That LNs would inhibit LNs makes sense when the goal is maximizing PN output of the active glomerulus: By inhibiting the LNs of rival glomeruli, the active glomerulus reduces the amount of inhibition directed at itself.
Similarly, that LNs would inhibit PNs makes sense if the goal is to reduce the PN output of rival glomeruli.

\paragraph*{Octopamine's effects on different neuron types}
Octopamine increases the responsivity of a neuron to incoming signals. 
It is unclear how or whether octopamine affects various neuron types (ie RNs, PNs, LNs, KCs).
Calibration of the Network Model's AL behavior, and tuning of  KC behavior to enable learning, indicate that octopamine needs to target RNs and LNs, but not PNs, KCs, or ENs.
Logical arguments support these findings:

%{\bf{RNs:}}
RNs: Because RNs initially receive the odor signal, these are logical neurons to stimulate with octopamine, because it sharpens their response to the exact signature being trained, which in turn sharpen the AL's output code for that odor.

%{\bf{LNs:}} 
LNs: LNs have the dual roles of inhibiting rival glomeruli and limiting overall PN output in the AL. 
For the first role, increased LN response to RNs will tend to sharpen AL response to the trained odor, by accentuating inhibition of rival glomeruli PNs.
For the second role, increased LN activity mitigates the risk that increased RN activity (due to octopamine) might blow up the overall PN output of the AL.

%{\bf{PNs:}}
PNs: Network Model simulations suggest that PNs should receive little or no octopamine stimulation.
While increasing PN responsivity would benefit RN-induced sharpening of the trained odor's signature, there are three downsides.
First, RN input to PNs is intrinsically noisy, so higher PN responsivity amplifies noise as well as signal.
Second, since PNs respond to LNs, higher PN activity tends to reduce the impact of LN inhibition, and thus reduces the inhibition-induced sharpening of the AL odor response caused by octopamine.
Third, increasing PN responsivity can have an outsize effect on overall PN firing rates, ie it is a  ``high-gain" knob and therefore risky.

%{\bf{KCs:}} 
KCs: Network Model simulations indicate that direct octopamine stimulation of KCs greatly reduces sparseness in the MB, which can be disastrous to learning.
Thus we expect that octopamine has no, or only slight, direct stimulative effect on KCs.

%------------------------------------------------------------------------------------------------------------------------------------------------------------------------

\subsection{Noise filtering role of the sparse, high-dimensional stage}~\\
Projection from a dense, low-dimensional coding space (eg the AL) to a sparse, high-dimensional coding space (eg KCs in the MB) is a widespread motif of biological neural systems, with size shifts are routinely on the order of 20x to 100x \cite{ganguli2012, babadi,litwinKumarHarris2017}.
The reasons for this pattern are not fully understood.
Some proposed reasons include information capacity, long-range brain communication, and reduced training data needs \cite{ganguli2012}, as well as improved discrimination ability \cite{peng2017, litwinKumarHarris2017}.

Network Model experiments bring to light another, central, role of sparseness: 
It acts as a robust noise filter, to protect the Hebbian growth process from amplifying upstream noise to out-of-control levels.
Though noise may be useful (or unavoidable) in upstream networks such as the AL, noise that reaches the neurons on both sides of a synaptic connection will be amplified by Hebbian growth during learning, swamping the system's downstream neurons (eg ENs) with noise.
However, the ``fire together, wire together" principle of Hebbian learning is an AND gate. 
Thus it suffices to remove noise from just one of the two connected neurons to prevent synaptic growth.
Sparsity does precisely this. 

Network Model experiments show that sparseness in the MB ensures that noise does not get amplified during training, so that post-training EN spontaneous firing rates and EN control odor responses remain unchanged.
Conversely, when KC response is not sufficiently sparse, any training leads rapidly to noisy EN spontaneous response levels and amplified EN responses to control odor.
This implies that the noise filtering induced by MB sparseness is necessary for a workable Hebbian learning mechanism.
This finding agrees with \cite{peng2017}, where experiments with a computational MB model indicated that sparse MB response gave stronger built-in discriminating ability. %(though without the context of Hebbian AND-gate dynamics given here). 

Setting aside the particular demands of Hebbian plasticity, robust  noise filtering may be a core function of sparse, high-dimensional stages within any network cascade where noise accumulates due to (beneficial) use in upstream stages.

%------------------------------------------------------------------------------------------------------------------------------------------------------------------------

\subsection{Roles of octopamine}~\\
The levels of octopamine stimulation in our Network Model were calibrated to \textit{in vivo} data on PN responses to octopamine. Thus, the simulations give insights into downstream effects of octopamine on plasticity, KC responses, EN responses, and Hebbian learning.\\

\paragraph*{Accelerant}
Moths can learn to respond to new odors remarkably quickly, in just a few exposures. 
Simulations indicates that while Hebbian growth can occur without octopamine, it is so slow that actionable learning, ie in terms of amplified EN responses, does not occur. 

This implies that octopamine, through its stimulative effect, acts as a powerful accelerant to learning. 
Perhaps it is a mechanism that allows the moth to work around intrinsic organic constraints on Hebbian growth of new synapses, constraints which would otherwise restrict the moth to an unacceptably slow learning rate. 
To the degree that octopamine enabled a moth to learn more quickly, with fewer training samples, it would clearly be highly adaptive.\\

\paragraph*{Active learning}
Simulations indicate that octopamine strongly stimulates the EN response to even an unfamiliar odor. 
Since octopamine is delivered as a reward, this has a beneficial effect in the context of reinforcement learning \cite{suttonBarto}, with the moth as the learning agent. 
An agent (the moth) can in some cases learn more quickly when it has choice as to the sequence of training samples (Active Learning \cite{settles}).

In particular, when a certain class of training sample is relatively rare, it benefits the agent to actively seek out more samples of that class \cite{attenberg}.
Octopamine enforces high EN response to a reinforced odor, ensuring that ENs will consistently exceed their ``take action" threshold during training. 
If the action is to ``approach", the moth is more likely to again encounter the odor, thus reaping the benefits predicted by Active Learning theory. 
This advantage applies in the context of positively-reinforced odors.

In the case of aversive learning, the high EN responses to unfamiliar but objectionable odors, due to dopamine, would cause the moth to preferentially avoid further examples of the odor. 
This would slow learning of aversive responses (a drawback), but would also minimize the moth's exposure to bad odors (danger avoidance, a benefit).\\ 
 
\paragraph*{Exploration of optimization space}
A limitation of Hebbian growth is that it can only reinforce what already exists.  
That is, it only strengthens channels that are transmitting signals deemed (by association) relevant to the stimulus being reinforced.  Absent a mechanism like octopamine, this constrains growth to channels that are already active. 
Simulations indicate that octopamine induces much broader activity, both upstream from and within the plastic layer,  thus activating new transmitting channels. 
This allows the system to strengthen, and bring permanently online, synaptic connections that were formerly silent.
This expands the solution space the system can explore during learning.
This function may be particularly important given the constraint of sparsity placed on odor codes in the MB.\\

\paragraph*{Injury compensation}
There is evidence that many forms of injury to neurons result in dropped spikes and thus lower firing rates in response to odors \cite{maiaReactionTime}.
This injury-induced drop in the signals reaching the ENs could induce behavioral consequences, by lowering EN responses to below key behavioral action thresholds. 
 In this case, the generalized stimulation induced by octopamine might compensate for the reduced signal strength sufficiently to lift the EN responses back above threshold during training. 
This in turn would allow Hebbian growth to strengthen the synaptic connections to those ENs, such that even the reduced signals from the still-injured upstream network would  be sufficient to push EN response above the action threshold.  
This mechanism (octopamine stimulation plus Hebbian synaptic growth) would allow the moth to regain behavioral function lost due to damage to the upstream neural system.

%------------------------------------------------------------------------------------------------------------------------------------------------------------------------
 
\subsection{The value of randomness}~\\
The principle of randomness permeates the moth olfactory network, for example in random neural connection maps \cite{caron2016} and in highly variable performance characteristics of chemical receptors in the antennae \cite{olsenWilson2010}. 
A key result from Network Model experiments is that a biologically-based neural net permeated with the random principle can robustly learn. 
This is in marked contrast to engineered computers, where poorly-spec'ed components are a liability.
Particular benefits of randomness include:\\

\paragraph*{Random KC$\rightarrow$EN connections} 
These guarantee that projections onto ENs are incoherent relative to whatever low-dimensional manifold in KC-space, which ensures (by the Johnson-Lindenstrauss lemma) that the EN readouts will preserve distinctions between elements in KC-space \cite{ganguli2012}.\\

\paragraph*{Variable gaba-sensitivity in glomeruli} 
 This increases the range of sensitivity to odor concentration, because some glomeruli simply don't turn off when a different, stronger odor tries to mask them through lateral inhibition (LNs) \cite{hong2015}.\\

\paragraph*{Variable sensitivity of antennae receptors} 
This gives a natural sensitivity to concentration, as progressively stronger odor will progressively activate the less-sensitive receptors, increasing the total RN input to glomeruli.\\

\paragraph*{Resistance to injury}
A randomly-connected network is more robust to damage. 
When exact connection maps and strengths are not required in the first place, damage has less impact on the fundamental nature of the system. % \newline

Most importantly, randomness (of connections and component properties) is evolutionarily cheap, easy, and available. 
So perhaps the core benefit of randomness to the moth olfactory network is that it works at all.

%---------------------------------------------------------------------------------------------------------------------------------------------------------------------

\subsection{The value of noise}~\\
Noise in biological neural systems is believed to have value, for example by encoding probability distributions and enabling neural Baysian estimations of posteriors \cite{maPougetBayesianInference}.
In addition, injection of noise while training engineered NNs can improve trained accuracy \cite{guozhong}.
Simulations indicate that in the AL-MB, noise has potential benefits, coupled with caveats.  
Figure \ref{kcConsistency} shows how noise in the AL adds an extra dimension to MB odor encoding, increasing the granularity of its odor responses.

The MB responds to odors in two ways: (i) by the number of KCs that are responsive, and (ii) by the reliability (eg from 10\% to 100\%) of their responses. 
This can be seen in the effect of octopamine on KC odor response, Fig \ref{kcConsistency} (B).  
Octopamine boosts MB odor response by increasing the number of active KCs (horizontal shift in response curves) and also by increasing the reliability of responsive KCs (vertical shift in responsivity curves).  
Both these shifts represent a stronger MB response and translate into stronger EN response. 

Taken together, they provide a finer granularity of the response range, versus the binary response of a noise-free system.  
Looked at another way,  the MB response to noisy inputs from the AL is a concrete example of a mechanism used by a neural system to translate the probability distributions encoded by noisy neurons into actionable signals with high dynamic range and granularity. 

A caveat is that noise in the AL-MB must be confined to the AL, i.e. upstream from the encoding layer, in order to protect the readout neurons and Hebbian learning mechanism from noise. 
The system's success depends on robust noise filtering at the MB layer, via global inhibition from the LH.  
So the three-stage architecture consisting of: %\newline

Noisy pre-amplifier layer $\rightarrow$ Sparse noise-reduction layer $\rightarrow$ Decision layer \newline
is an interdependent system well-suited to nuanced decision-making.
  
Given a layer (the MB) that effectively protects the decision neurons from upstream noise, the system is also potentially robust to noisy stimuli. 
In the neural net context, input samples (i.e. inputs to the feature-reading layer) can be thought of as a \textit{de facto} ``first layer" of the neural net.  
A system that is robust to upstream noise may also be naturally robust to noisy inputs, a further potential advantage of judicially-placed sparse layers.

\subsection{Applications to Machine Learning}~\\
The model and simulations in this paper characterize key features of the AL-MB system which might usefully be ported to machine learning algorithms.  
These features include: 
Generalized stimulation during training; Hebbian growth; sparse layers to control plastic connections and filter noise; and noisy initial layers.
Advantages of this biological toolkit include:\\ 

\paragraph*{Fast learning} Moths can reliably learn a new odor in less than 10 exposures.
Biological brains in general can learn given few training samples.
This contrasts by orders of magnitude with the voracious demands of DNNs, where assembling sufficient training data can be a serious chokepoint in deployment.\\

\paragraph*{Robustness to noise}
The sparse layer in the AL-MB acts as an effective noise filter, protecting the readout neurons from a noisy upstream layer (the AL).  
Since the system is designed to accommodate upstream noise, it is possible that it can also readily accommodate noisy input samples.  
NNs have a troublesome property, that input-output score functions are not locally continuous \cite{szegedy}.  
Biological neural nets seem to avoid this particular fault.  
The noisy layer $\rightarrow$ sparse layer motif may be one reason for this.  
It may thus be a useful motif to apply in ML architectures. \\ 
%We are currently pursuing these ML applications, by adapting the AL-MB system to various ML tasks, and by applying the key structures and mechanisms described here to more complex NN architectures.
 
\paragraph*{Novel training mechanism}
Hebbian growth, combined with generalized stimulation via octopamine, is (in the context of ML) a novel mechanism to explore a solution space and train a classifier. 
In particular, it works on a fundamentally different principle than backpropagation algorithms: 
It does not minimize a loss function via gradient descent; rather, it selectively strengthens only those connections that are transmitting meaningful signals. 
The addition of a sparse layer acts to control and focus learning, by leveraging the AND-function nature of Hebbian growth.\\

\paragraph*{Biological plausibility}
One characteristic (not criticism) of backprop optimization is that it is biologically implausible, since it requires a neuron to have more than local knowledge of the system. 
The search for neural network architectures (for example with recurrent connections to transport non-local information) and variants of backprop which are biologically plausible, and which thus might narrow the gap between biological and engineered NNs, is currently an area of interest, especially in the context of DNNs \cite{bengioNIPS2015}.  
This paper demonstrates that the triad of octopamine stimulation + Hebbian growth + sparse layers can efficiently train a NN, and is thus a possible candidate to address the biological plausibility gap. \\

% \clearpage
 
%%%%%%%%%%%%%%%%%%%%%%%%%%%%%%%%%%%%%%%%%%%%%
%%%%%%%%%%%%                   Materials and Methods                         %%%%%%%%%
%%%%%%%%%%%%%%%%%%%%%%%%%%%%%%%%%%%%%%%%%%%%%

\section{Materials and Methods}~\\
In this section, we describe the biological moth olfactory network, as well as our Network Model. 
We also provide a Glossary, and describe the \textit{in vivo} data used for model calibration.

\subsection{Moth olfactory system overview}~\\ \label{systemOverview}
The parts of the AL-MB implicated in learning are organized as a feed-forward cascade of five distinct networks, as well as a reward mechanism \cite{martin2011, kvello2009}.
Figure \ref{schematicPlusTimecourses} gives a system schematic along with typical firing rate (FR) timecourses (from simulation) for neurons in each network.

% These networks are:
\begin{enumerate}

\item Antennae. Roughly 30,000 noisy chemical receptors detect odor and send signals to the Antenna Lobe \cite{masse2009}.
\item Antenna Lobe (AL). Contains roughly 60 units (glomeruli), each focused on a single odor feature \cite{martin2011}.
The AL essentially acts as a pre-amp, boosting faint signals and denoising the antennae inputs \cite{bhandawat2007}.
 AL neurons are noisy \cite{galizia2014}.
\item Lateral Horn (LH). Though not fully understood, one key function is global inhibition of the Mushroom Body to enforce sparseness \cite{bazhenovStopfer2010}.
\item Mushroom Body (MB), here synonymous with the Kenyon Cells (KCs). About 4000 KCs are located in the calyx of the Mushroom Body (MB). These fire sparsely and are believed to encode odor signatures \cite{perisse2013, campbellMushroomBody, honeggerTurner2011}.
\item Extrinsic Neurons (ENs), numbering $\sim$10's, located downstream from the KCs. These are believed to be ``readout neurons" that interpret the KC codes and convey actionable messages (such as ``fly upwind") \cite{campbell2013, hige2015}.
\item Reward Mechanism. A large neuron sprays octopamine globally over the AL and MB, in response to reward, such as sugar at the proboscis. Learning does not occur without this octopamine input \cite{hammer1995, hammerMenzel1998}.
\item Inter-network connections: In the AL-MB these are strictly feed-forward, either excitatory or inhibitory. In particular, Antennae$\rightarrow$AL, AL$\rightarrow$LH, KCs$\rightarrow$ENs are all excitatory. LH$\rightarrow$KCs is inhibitory. AL$\rightarrow$KCs have both excitatory and inhibitory channels.
\item Plasticity: The connections into the KCs (AL$\rightarrow$KCs) and out of the KCs (KCs$\rightarrow$ENs) are known to be plastic during learning \cite{cassenaer, masse2009}. 
The AL is not plastic.
\end{enumerate}

%-----------------------------------------------------------------------------------------------------------------------------------------------------------------

\subsection{Glossary} \label{glossary}
\paragraph*{Antenna lobe (AL)} A collection of neurons innervated by odor receptors in the antennae. 
It sends signals to the mushroom body via projection neurons.
Connections in the AL are not plastic.%  \newline  \newline
\paragraph*{Mushroom body (MB)} A collection of neurons (Kenyon cells - KCs) downstream from the antenna lobe. 
The MB is believed to store odor codes that serve as a memory, allowing the moth to recognize odors.
Connections in the MB are plastic. % \newline \newline
\paragraph*{Lateral horn (LH)} A collection of neurons which receives input from the AL and sends inhibitory output to the MB. One of its roles is to enforce sparse firing in MB neurons. %\newline \newline
\paragraph*{Receptor neuron (RN)} These neurons respond to odors (volatiles) at the antennae and stimulate the antenna lobe. RNs respond to different, distinct odors. %\newline
\paragraph*{Glomerulus} The antenna lobe is divided into about 60 glomeruli, each of which is a self-contained collection of neurons (projection and lateral), innervated by RNs that respond to particular odors.%\newline \newline
\paragraph*{Projection neuron (PN)}  Each glomerulus contains projection neurons, whose output innervates the KCs and also the lateral horn, but not other glomeruli in the AL, ie they are feed-forward only. 
Most PNs start in one glomerulus and are excitatory. A few PNs arborize in several glomeruli and are inhibitory (we refer to inhibitory PNs as ``QNs"). Each glomerulus  initiates about five PNs.%\newline \newline
\paragraph*{Lateral neuron (LN)} Each glomerulus contains lateral neurons, which innervate other glomeruli in the AL. 
LNs are inhibitory. 
One function is competitive inhibition among glomeruli.
Another function is gain control, ie boosting low signals and damping high signals. %\newline \newline
\paragraph*{Kenyon cell (KC)} Neurons in the calyx of the MB. These have very low FRs, and tend to respond to particular combinations of PNs. 
KCs respond sparsely to a given odor.
There are about 4000 KCs, ie a two-orders-of-magnitude increase over the number of glomeruli. 
Each KC synapses with about ten PNs.
Connections into and out of KCs are plastic.%\newline \newline
\paragraph*{Extrinsic neuron (EN)} A small number of neurons downstream from the KCs. ENs are thought to be ``readout" neurons. They interpret the odor codes of the KCs, deciding to eg ``ignore", ``approach", or ``avoid".%\newline
\paragraph*{Firing rate (FR)} The number of spikes/second at which a neuron fires. Typically FRs are counted using a window (eg 500 ms). 
The moth's response to odor stimulations is episodic, with FR spikes in FR and rapid return to spontaneous FRs.
Neurons respond to relative changes in FR, rather than to raw magnitude changes. A neuron's relative change in FR is scaled by its spontaneous FR (see section \ref{firingRateMeasure} below).%\newline \newline
\paragraph*{Octopamine} A neuromodulator which stimulates neural firing. The moth spritzes octopamine on both the AL and MB in response to sugar, as a feedback reward mechanism. 
Dopamine has a similar stimulating effect on both AL and MB, but it reinforces adverse rather than positive events.  \newline

%------------------------------------------------------------------------------------------------------------------------------------------------------------
%------------------------------------------------------------------------------------------------------------------------------------------------------------

\subsection{Component networks and their Network Model representations}~\\ \label{componentNetworks}
This subsection offers a more detailed discussion of the constituent networks in the biological AL-MB, and details about how they are modeled in our Network Model.

\paragraph*{Antennae and receptor neurons}
The Antennae receptors, activated by chemical molecules in the air, send excitatory signals to Receptor Neurons (RNs) in the AL. 
Several thousand antennae converge onto 60 units (glomeruli) in the AL \cite{nagel2011}. 
All the receptors for a given atomic volatile converge onto the same glomerulus in the AL, so the glomeruli each have distinct odor response profiles \cite{deisig2006}.
Since natural odors  are a blend of atomic volatiles, a natural odor stimulates several units within the AL \cite{riffellPnas2009}.  

Our model does not explicitly include antennae. 
Rather, the first layer of the model consists of  the RNs entering the glomeruli.  
Though $\sim$500 RNs feed a given glomerulus, the model assumes one RN. 
The benefit of many RNs converging appears to be noise reduction through averaging \cite{olsenWilson2010}.
This can be simulated by one RN with a smaller noise envelope.
Each glomerulus' RN has a spontaneous FR and is excited, according to random weights, by odor stimuli.

\paragraph*{Antenna lobe and projection neurons} 
The AL is fairly well characterized in both structure and dynamics, with a few important gaps. 
Moths and flies are similar enough that findings in flies (\textit{Drosophila}) can generally be transferred to the moth  (in contrast, locusts and honeybees are more complex and findings in these insects do not safely transfer) \cite{riffellCurrBio2009}.  

\begin{figure}[t]
%\begin{adjustwidth} {-60mm}{0mm}
\begin{center}
\includegraphics {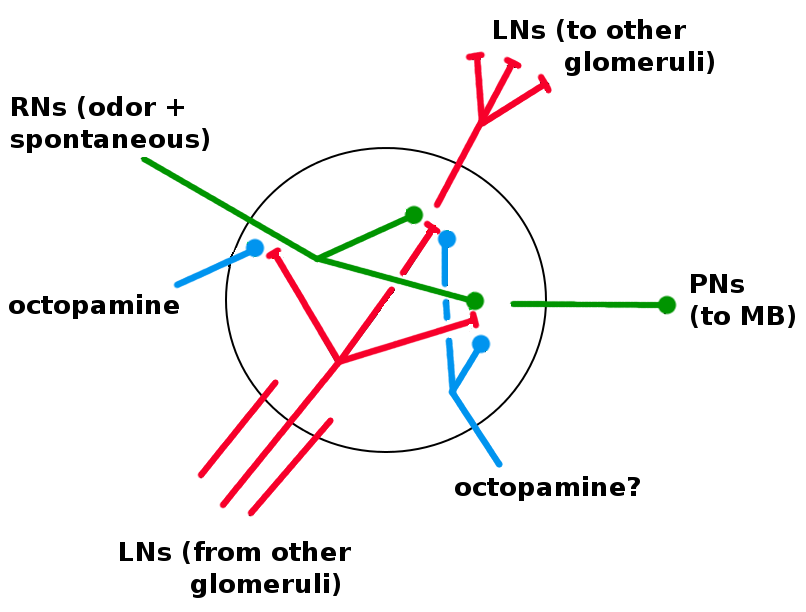}  
\caption{{\bf Schematic of AL glomeruli.} 
Detail of neural connections within a glomerulus. 
Red = inhibitory, green = excitatory, blue = increases responsiveness.
RNs enter from the antennae. LNs enter from other glomeruli; one full LN is shown.
It is not known if octopamine modulates LNs and PNs (see section \ref{discussionPredictions}).  }  
\label{glomSchematic}
\end{center}
%\end{adjustwidth}
\end{figure}

The AL contains about 60 glomeruli, each a distinct unit which receives RN input and projects to the KCs via excitatory PNs.
The same PN signal also projects to the LH \cite{bazhenovStopfer2010}.  
The AL, unique among the networks, has inhibitory lateral neurons (LNs) \cite{wilson2005}, the only neurons that are not strictly feed-forward.
(There is some evidence of excitatory LNs, eg \cite{olsen2008}; the Network Model excludes this possibility.)
The LNs act as a gain control on the AL, and also allow odors to mask each other by inhibiting other glomeruli's RNs \cite{olsenWilson2008, hong2015}.  
It is not known whether LNs also inhibit PNs and LNs.
Based on calibrations to \textit{in vivo} data, in Network Model LNs inhibit all neuron types (cf section \ref{discussionPredictions}).
Thus each glomerulus contains dendrites (ie outputs) for PNs and LNs, and axons (ie inputs) for RNs and LNs, as shown in Figure \ref{glomSchematic}.

Each glomerulus does the following:  
Receives RN input from the antennae receptors upstream; 
inhibits other glomeruli within the AL via LNs;
and sends excitatory signals downstream via Projection Neurons (PNs).

In general, each PN is innervated in a single glomerulus. 
In moths, there are $\sim$5 PNs rooted in each glomerulus (60 glomeruli, $\sim$300 PNs).
The Network Model assumes all PNs from a given glomerulus carry the same signal (because they share the same glomerulus and therefore inputs, and perhaps also because of ephaptic binding) \cite{sjoholm2006}.  

Glomeruli also initiate pooled Inhibitory Projection Neurons (QNs) that send inhibitory signals downstream to the KCs. 

The AL contains a powerful macro-glomerulal complex (MGC), which processes pheromone.
Because pheromone response has fundamentally different dynamics than food odor response \cite{jefferisPotter2007}, the model ignores it. 
Only the glomeruli associated with non-pheromone (food) odors are modeled.

Connections in the AL are not plastic with long-term persistence \cite{dacksRiffell2009}. 
While some evidence of short-term plasticity exists, the Network Model ignores this option.

\paragraph*{Lateral horn}
The LH receives input from the PNs. 
It then sends an inhibitory signal to the KCs. 
This inhibition from the LH appears to ensure that the KCs fire very sparsely and thus act as coincidence detectors for signals from the AL \cite{sjoholm2006, lin2014, gruntman}.  

The LH is also suspected of containing a parallel system for processing certain intrinsically-known odors in short-cut fashion (labeled lines) \cite{luo2006}.
Since this parallel system is (by definition) not involved with learning, the Network Model ignores it. 
The LH is modeled solely as a simple sparsifying inhibition on the KCs.

(Note: The locust and honeybee, which have more complex olfactory systems and different use-cases in terms of odor processing, have a time-oscillating depolarization mechanism (local potential fields, LPF) \cite{perezOrive} which serves a similar purpose to LH inhibition in the moth. LPF oscillations are absent in the moth \cite{martin2011}.) 

\paragraph*{Mushroom body and Kenyon cells}
The KCs ($\sim$4000) in the MB are believed to encode odor memories in a high-dimensional, sparse space \cite{turner2008}.  
Odors with no meaning to the moth still have non-zero codes in the KCs.

KCs receive excitatory input from the PNs and inhibitory input from QNs, both of which vary greatly between KCs, since each KC is innervated by only $\sim$10 PNs \cite{martin2011}.
The connection map appears to be random \cite{caron2013}.  
The KCs also receive generalized damping inhibition from the LH. 
(There is some evidence in drosophila of an MB$\rightarrow$MB global inhibitory neuron \cite{lin2014}, with the same essential effect as LH inhibition; the Network Model excludes this possibility.)
KCs fire very sparsely, generally respond to only a single odor, and are silent absent that odor \cite{honeggerTurner2011}.  
KCs are treated as noise-free.
Their output is an excitatory signal sent to the extrinsic neurons (ENs) \cite{campbell2013}.  

In addition to olfactory input, the KCs receive input signals from other parts of the moth (eg hearing) \cite{sjoholm2006}.
Because the Network Model targets olfactory learning, it ignores these other inputs and uses a reduced number of KCs ($\sim$2000 instead of $\sim$4000).

The synaptic connections in the MB (PNs$\rightarrow$KCs,  QNs$\rightarrow$KCs, and KCs$\rightarrow$ENs) are plastic, ie they can be modified  during training \cite{menzelManz}.  
The generalized inhibition from LH$\rightarrow$KCs is modeled as non-plastic (actual physiology is not known).

\paragraph*{Extrinsic neurons}
Though located in the lobes of the MB, here ENs are not considered part of the MB, which is taken to be synonymous with the KCs.
ENs are few in number compared to the KCs ($\sim$10s ) \cite{campbell2013, hige2015}.
They are believed to be ``readout" neurons, that interpret the KC codes as actionable signals (eg ``approach", ``avoid") \cite{masse2009}.
We assume that ENs trigger actions when their output FRs exceed some threshold.

We define Learning as: Permanently boosting EN responses beyond their naive (untrained) level, so that EN responses to reinforced stimuli can consistently exceed an action-triggering threshold.
This is tantamount to modifying the moth's behavior.

\paragraph*{Octopamine (reward circuit)}
A large neuron delivers octopamine to the entire AL and MB, in response to positive stimuli, eg sugar at the proboscis.
It acts as a reward feedback to the system.
A similar neuron delivers dopamine to the AL and MB in response to negative stimuli, and acts as an aversive feedback signal \cite{dacksRiffell2009}.
Learning does not occur without octopamine (or dopamine)  \cite{hammerMenzel1998}.  

Despite their opposite reward values, both octopamine and dopamine act in the same way when sprayed on a neuron: 
They increase the neuron's general tendency to fire \cite{riffell2013}.
% see section~%section_1.2.9    % add period
In the Network Model this effect is modeled as making a neuron more responsive to excitatory inputs (eg from odors and RNs) and less responsive to inhibitory inputs (eg from LNs).
Details of octopamine's effects if any on various neural types are not well-characterized.
In the Network Model octopamine directly affects RNs and LNs but not PNs in the AL (cf section \ref{discussionPredictions}); 
has no direct effect on KCs or ENs (though there are strong indirect effects); 
and has no effect on the LH inhibitory signal.

It is unclear whether octopamine delivery to both the MB and AL is necessary and sufficient for learning \cite{hammerMenzel1998, dacksRiffell2009}.
The Network Model assumes that octopamine controls an ``on/off" switch for Hebbian growth, ie there is no plasticity in the MB (and therefore no learning) without octopamine.
%------------------------------------------------------------------------------------------------------------------------------------------------------------
%------------------------------------------------------------------------------------------------------------------------------------------------------------

\subsection{Network Model description}~\\
This section describes our Network Model  in detail. 
It covers the firing rate measure used to compare model output to \textit{in vivo} data; model dynamics; plasticity and other details;  model parameters; and moth generation.
All coding was done in Matlab.
Computer code for this paper will be found at:\newline 
\url{https://github.com/charlesDelahunt/SmartAsABug} % \newline \newline 

\subsection{Firing rate measure}~\\  \label{firingRateMeasure}
To compare PN firing rate statistics from \textit{in vivo} experiments and from Network Model simulations (ie model calibration), we use a measure of firing rate (FR) based on Mahalanobis distance, similar to the measure $\frac{DF}{F}$ common in the literature \cite{campbell2013, hong2015, turner2008, silbering2007}. 
The premise is that neurons downstream respond to a $\pm$1 std change in FRs equally (modulo different connection weights), independent of the sometimes large (up to 40x) magnitude differences in the raw spontaneous FRs of different neurons. 
The FR measure is defined as follows:
\begin{enumerate}
\item Each PN has a spontaneous firing rate (FR) with a gaussian noise envelope.
\item PNs with FR $<$ 1 spike/sec are ignored, on the assumption that such PNs represent artifacts of experiment (also, the gaussian noise assumption fails).
About 10\% of PNs in experimental data fall in this category.
\item Output FR activity of PNs is measured as $M(t)$ = distance from mean spontaneous FR, in units of time-varying std dev of spontaneous FR (ie Mahalanobis distance):
Let \newline
$F(t)$ = raw firing rate (spikes per second). \newline
$S(t)$ = spontaneous firing rate (no odor).\newline
$\mu S(t)$ = moving average of S (no odor). \newline
$\bar{\mu}S(t)$ = smoothed estimate of the moving average $\mu S$, eg a quadratic or spline fit. \newline
$\sigma_{S}(t)$ = standard deviation of $S$, calculated using $S - \bar{\mu}S$ values within a moving window centered on $t$. \newline
$\sigma_{S}(t)$  and $\mu S(t)$ are typically steady absent octopamine, but are often strongly modulated by octopamine.\newline 
Then the measure of FR activity $M$ is:\newline
\begin{equation}
M(t) = \frac{F(t) - \bar{\mu}S(t)}{\sigma_{S}(t)}
\end{equation}
\item $M$ is related to the measure $\frac{DF}{F}$:\newline
$\frac{DF}{F}= \frac{\Delta F}{F} = \frac{F(t) -  \mu S}{\mu S}$, ie $\frac{DF}{F} $ is change in FR, normalized by spontaneous FR.  	
The key difference between $M$ and $\frac{DF}{F}$ is whether or how $\sigma_S$ is estimated, due to varying exigencies of experiment.
Our experimental data allow reasonable estimates of $\sigma_S$ and $\mu S$. 
Network Model simulations produce very good estimates, since computers are more amenable to repeated trials than live moths.

\end{enumerate}

%-------------------------------------------------------------------------------------------------------------------------------------------
%-------------------------------------------------------------------------------------------------------------------------------------------

\subsection{Model dynamics}~\\  \label{dynamics}
Our Network Model uses standard integrate-and-fire dynamics \cite{dayan2001},  evolved as stochastic differential equations \cite{higham2001}.\newline  
Let $x(t)$ = firing rate (FR) for a neuron. Then
\begin{equation}
 \tau \dfrac{dx}{dt} =  -x + s (\Sigma {\bf{w}}_{i} {\bf{u}}_{i} ) = -x + s(\bf{w}\cdot\bf{u}), \text{ where}
\end{equation}
$~~~${\bf{w}} = connection weights;\newline
$~~~$ {\bf{u}} = upstream neuron FRs; \newline
$~~~~$$s()$ is a sigmoid function or similar.\newline 

PN dynamics are given here as an example. 
Full model dynamics are given in Section \ref{fullDynamics}.
PNs are excitatory, and project forward from AL$\rightarrow$MB:
\begin{equation}
\tau \dfrac{d{\bf{P}} }{dt}  =  - {\bf{P}} + s( \widetilde{\bf{P} } ) +  {\bf{dW}}^{P} \text{ where}
\end{equation}
$~~$${\bf{W}}(t)$ = brownian motion process; \newline
$~~~$$\widetilde{\bf{P}} = -(1- \gamma o(t)M^{OP} ) \text{*} {M^{LP}} \text{*} {\bf{u}}^{L} + (1 + o(t)M^{OP}  ) \text{*} M^{RP} \text{*}{\bf{u}}^{R} $; \newline
$~~~$$M^{OP}$ = octopamine$\rightarrow$PN weight matrix (diagonal $nG \times nG$);  \newline
$~~~$$M^{LP}$ = LN$\rightarrow$PN weight matrix ($nG \times nG$ with $trM^{LP} = 0$);  \newline
$~~~$$M^{RP}$ = RN$\rightarrow$PN weight matrix (diagonal $nG \times nG$); \newline
$~~~$$o(t)$ indicates if octopamine is active ($o(t)$ = 1 during training, 0 otherwise).\newline
$~~~$${\bf{u}}^{L}$ = LN FRs, vector $nG \times 1$;  \newline
$~~~$${\bf{u}}^{R}$ = RN FRs ($nG \times 1$);  \newline 
$~~~$$\gamma$ = scaling factor for effects on inhibition.\newline

\paragraph*{Discretization}
The discretization uses Euler-Maruyama, a standard step-forward method for SDEs  \cite{higham2001}.\newline
Euler (ie noise-free): $x_{n + 1} = x_{n} + \Delta t f(x_{n} )$\newline
Euler-Maruyama:  $x_{n + 1} = x_{n} + \Delta t f(x_{n} ) + \epsilon \text{ randn(0,1)} \sqrt{\Delta t}$, where $\epsilon$ controls the noise intensity.  

\paragraph*{Convergence} 
Timestep $\Delta t$ was chosen such that noise-free E-M evolution gives the same timecourses as Runge-Kutta (4th order), via Matlab's ode45 function.
$\Delta t = 10$ mSec suffices to match E-M evolution to R-K in noise-free moths.
Values of $\Delta t \leq 20$ mSec gives equivalent simulations in moths with AL noise calibrated to match \textit{in vivo} data. 
Values of $\Delta t \geq 40$ mSec show differences in evolution outcomes given AL noise.

\paragraph*{Plasticity}
The model assumes a Hebbian mechanism for growth in synaptic connection weights \cite{hebb, cassenaer}. 
That is, the synaptic weight $w_{ab}$ between two neurons $a$ and $b$ increases proportionally to the product of their firing rates (``fire together, wire together''):  
 $\Delta w_{ab}(t) \propto f_a(t) f_b(t). $ \newline
Thus, synaptic plasticity is defined by: 
\begin{equation} \label{hebbianGrowthEqn}
\Delta w_{ab}(t) = \gamma f_a(t) f_b(t), \text{ where } \gamma \text{ is a growth parameter. } 
\end{equation}

There are two layers of plastic synaptic weights, pre- and post-MB:  AL$\rightarrow$MB ($M^{P,K},M^{Q,K}$), and MB$\rightarrow$ENs ($M^{K,E}$) .
Learning rate parameters of the Network Model were calibrated to match experimental effects of octopamine on PN firing rates and known moth learning speed (eg 4 - 8 trials to induce behavior modification) \cite{riffell2013}.
The Network Model does not decay unused synaptic weights.
Training does not alter octopamine delivery strength matrices ($M^{O,\text{*}}$). 
That is, the neuromodulator channels are not plastic (unlike, for example, the case in \cite{grant}).  

\paragraph*{Odor and octopamine injections}
Odors and octopamine are modeled as hamming windows.
The smooth leading and trailing edges ensures low stiffness of the dynamic ODEs, and allows a 10 mSec timestep to give accurate evolution of the SDEs in simulations.

\paragraph*{Training}
Training on an odor consists of simultaneously applying stimulations of the odor, injecting octopamine, and ``switching on" Hebbian growth.
Training with 5 to 10 odor stimulations typically produces behavior change in live moths. 

%----------------------------------------------------------------------------------------------------------------------------------------------------
%-------------------------------------------------------------------------------------------------------------------------------------------

\subsection{Model parameters}~\\   \label{parameterChoices}
There is a risk, when modeling a system, of adding too many free parameters in an effort to fit the system.
Fewer free parameters are better, for the sake of generality and to avoid overfitting. Conversely, we wish to reasonably match the physiological realities of the system.
Because the key goal of this paper was to demonstrate that a simple model, in terms of parameters and structure, can reproduce the learning behavior of the AL-MB, we made efforts to minimize the number of free parameters. 
For example, neuron-to-neuron connections in the model are defined by their distributions, ie two parameters each. 
These are (usually) distinct for different source-to-target pairs (eg LN$\rightarrow$RN,  LN$\rightarrow$LN, etc). \newline
Some mean and std dev parameters for distributions are shared among different neuron types (eg LNs, PNs, and QNs all share the same variance scaling parameter).  
\paragraph*{Parameter list} There are in total 47 free parameters: \\
$~~~$1. Structure:  5 (eg number of neurons in each network) \\
$~~~$2. Dynamics:  12 (noise: 2. decay tau and sigmoid: 3. Hebbian growth: 6. misc: 1). \\
$~~~$3. Spontaneous RN FRs:  3.  \\
$~~~$4. Connection matrices: 27  (to control non-zero connection ratios: 5; synaptic weights (eg $M^{P,K}, M^{R,P}$): means 12, std devs 4; octopamine weights (eg $M^{O,R}, M^{O,P}$): means 6, std devs 2).  
% $~~~$Total free params: 47

\paragraph*{Dynamics parameters}
The differential equations of all neuron types share the same decay rate, set to allow return to equilibrium in $\sim$1 second, consistent with \textit{in vivo} data.
Neurons also share parameters of the sigmoid function within the differential equation.
Noise added via the SDE model is controlled by a single parameter $\epsilon$, the same for all neuron types. 
It is determined by empirical constraint on $\frac{\sigma_{S} } { \mu S}$, as shown in column 2 of Figure \ref{comparisonPlots}.

\paragraph*{Connection matrix generation}
Connection weight matrices (eg $M^{P,K}$ etc) are generated in a standard way, from Gaussian distributions with std dev $\sigma$ defined proportional to the mean $\mu$, using a scaling factor $v$: \newline
$M^{*,*}\sim N(\mu_{c}, \sigma_{c}^{2})$ where $\mu_{c}$ depends on the neuron types being connected, and $\sigma_{c} = v \mu_{c}$. Many connection types typically share the same $v$.

A special feature of the AL is that all the neurons in a given glomerulus share a common environment. 
For example, all the neurons, of whatever type, in  glomerulus \textit{A} will share the same strong (or weak) LN axon from glomerulus \textit{B}.
Thus, the RN, LN, and PNs in a given glomerulus are all correlated.
In addition, neuron types are correlated. 
To model this dual set of correlations, connection matrices in the AL are generated as follows.
As an example, consider LN connection matrices in the AL:
\begin{enumerate}
\item A glomerulus-glomerulus connection matrix $M^{L,G}$ is created, which defines LN arborization at the glomerular level.
\item This connection matrix is multiplied by a neural type-specific value to give $M^{L,P}$,$M^{L,L}$, and $M^{L,R}$ connection matrices. This is particularly important when tuning the various inhibitory effects of LNs on RNs, PNs (QNs), and LNs.
\item Sensitivity to GABA: A separate variance factor determines glomerular sensitivity to GABA (ie sensitivity to inhibition). This is tuned to match data in the literature \cite{hong2015}, and applies to LN-to-PN(QN) (ie $M^{L,P}$) connections only.
\end{enumerate}
The goal of this two-stage approach is to enforce two types of similarity found in the AL: (i) Connections to all neurons within a single glomerulus are correlated; and(ii) connections to all neurons of a certain type (LN, PN, RN) are correlated.

Due to constraints of the biological architecture there are many zero connections. 
For example,  about 85\% of entries in the AL$\rightarrow$MB weight matrix are zero because MB neurons connect to only $\sim$10 projection neurons \cite{caron2013}.  
All MB$\rightarrow$EN weights are set equal at the start of training. 
Training leads rapidly to non-uniform distributions as inactive connections decay and active connections strengthen.

\paragraph*{RN spontaneous firing rates}
RNs in the glomeruli of the AL have noisy spontaneous firing rates \cite{bhandawat2007}.
The Network Model simulates this by assigning spontaneous firing rates to RNs.
These spontaneous firing rates are drawn from a gamma distribution plus a bias: \newline
$\gamma(x| \alpha , \beta, b ) = b + \frac{\beta ^{\alpha} }{ \Gamma (\alpha)} x^{\alpha - 1} e^{- \beta x} $, where $\alpha , \beta $ are shape and rate parameters, and $\Gamma ( \cdot )$ is the Gamma function. \newline
This can be thought of as a source of energy  injected into the system, at the furthest upstream point (absent odor).
Other energy sources are odor signals and octopamine.
The spontaneous firing rates of all other neurons in the Network Model are the result of their integrate-and-fire dynamics responding as RN spontaneous FRs propagate through the system.

%-------------------------------------------------------------------------------------------------------------------------------------------
%-------------------------------------------------------------------------------------------------------------------------------------------

\subsection{Discrepancies between biology and model}~\\  \label{discrepancies}
There are some known discrepancies between our Network Model and the moth AL-MB. These are listed below.

\paragraph*{Connection weight distributions} 
This model version uses gaussian distributions to generate initial connection weights. 
However, moths used in live experiments are older and thus presumably have modified PN$\rightarrow$KC and KC$\rightarrow$EN connection weights. 
If this modification was strong, we might expect the connection weight distributions to tend towards a scale-free rather than gaussian distribution \cite{barabasi1999}.
This represents an unknown discrepancy between structure parameters of the live moths used in experiments vs the model.

\paragraph*{Hebbian pruning}
The Network Model contains no pruning mechanism to offset, via decay, the Hebbian growth mechanism. Such pruning mechanisms are common in nature, so it is reasonable to suppose that one might exist in the AL-MB. 
The moth has inhibitory as well as excitatory feed-forward connections from AL to MB. In the Network Model, pruning is functionally replaced by Hebbian growth of QN$\rightarrow$KC inhibitory connections, which act to inhibit KCs and thus offset the growth of excitatory PN$\rightarrow$KC connections (this does not directly offset KC$\rightarrow$EN Hebbian growth). 
Thus omitting a separate Hebbian decay mechanism is a matter of convenience rather than a match to known biology.

\paragraph*{Non-olfactory input to KCs}
In addition to olfactory input, the KCs receive signals from other parts of the moth, eg hearing.
Because this model targets only olfactory learning, it ignores these other inputs to the KCs, and reduces the total number of KCs (from $\sim$4000 to $\sim$2000).

\paragraph*{Number of QNs}
There are believed to be about 3-6 QNs projecting from the AL to the MB.
This model sets their number at about 15.
The reason is that, absent a Hebbian pruning system in the model, the QNs function as the brake on runaway increases in KC responses due to Hebbian growth.
So the increased number of QNs is a compensation for the lack of a weight-decay system.

\paragraph*{Number of ENs}
This model version has only one EN, since its goal is to demonstrate simple learning. 
The moth itself possesses multiple ENs.

\paragraph*{LH inhibition}
The LH$\rightarrow$KC inhibitory mechanism used in this chapter is modeled as a time-invariant global signal, delivered equally to all KCs. This simplifies the model parameter space while retaining the essential functionality of the LH.
A more refined version of LH$\rightarrow$KC inhibition might vary in strength according to PN output, since the same PN signals that excite the KCs also excite the LH.
The actual dynamics of the AL$\rightarrow$LH$\rightarrow$KC linkage are not known, beyond the principle that inhibition from the LH sparsifies the KC codes and makes the individual KCs act as coincidence detectors.
% 
%------------------------------------------------------------------------------------------------------------------------------------------------------------
%------------------------------------------------------------------------------------------------------------------------------------------------------------

\subsection{\normalfont\bfseries\textit{in vivo} neural recordings data}~\\  \label{experimentalData}
Model parameters were calibrated by matching Network Model performance to \textit{in vivo} electrode readings from the ALs of live moths. 
The various performance metrics are described in Results.

\begin{figure}[t]
%\begin{adjustwidth} {-55mm}{0mm}
\centering
\includegraphics[width=120mm] {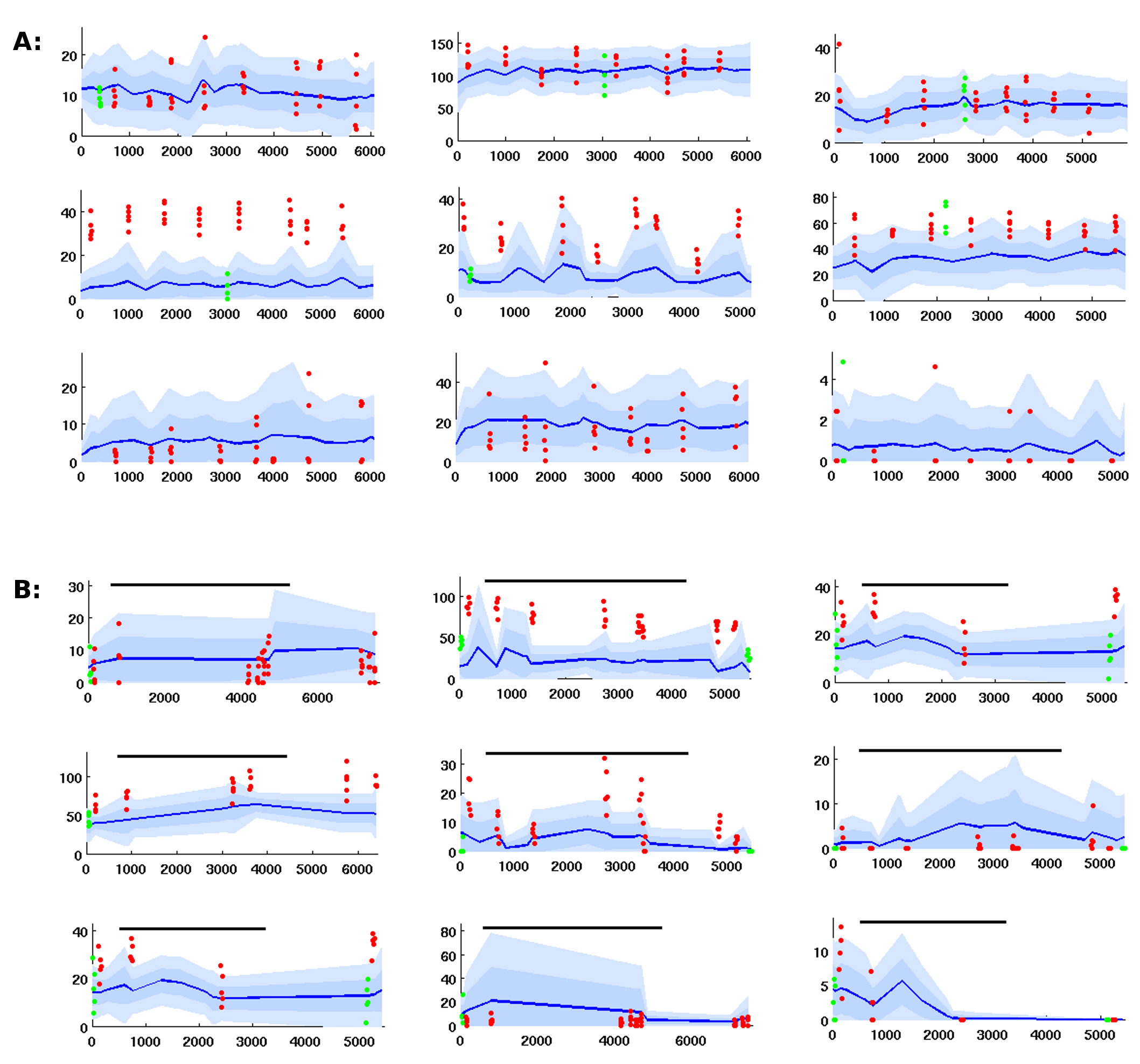}   
%\vspace*{-.2in}
\caption{ {\bf Time series of PN firing rates from \textit{in vivo} experiments.}
x-axis = time, y-axis = FR.
Blue lines = mean spontaneous rate, shaded regions = $\pm$1 and 2 std. 
Red dots are odor responses.
Green dots are response to control (mineral oil).  \newline
\textbf{A:} PN response, given odor plus coincident sugar reward, ie plus octopamine (time series for PNs with odor only are similar, but with less strong odor responses).
Top row: unresponsive to odor.
Middle row: excited response to odor.
Bottom row: inhibited response to odor. \newline
\textbf{B:} PNs with octopamine wash added in mid-experiment, then rinsed away (duration shown by black line).
Octopamine can alter (up, down, or not at all) the spontaneous FR and/or the odor response, so there are 9 possible modulation regimes.
This grid of timecourses shows a typical PN from each regime.
Top row: spontaneous FR in unaffected.
Middle row:  spontaneous FR is boosted.
Bottom row: spontaneous FR is inhibited.  
First column: odor response is unaffected.
Second column: odor response is boosted.
Third column: odor response is inhibited.       }
\label{pnTimecoursesJeff}
%\end{adjustwidth}
\end{figure}

Electrode data was collected by the lab of Prof Jeff Riffell (Dept of Biology, UW). It consists of timecourses of PN firing rates measured via electrode in the AL of live moths, during a variety of regimes including:

\begin{enumerate} 

\item Series of 0.2 sec odor stimulations delivered without octopamine. 
These experiments gave data re PN response to odor relative to PN spontaneous (baseline) FRs, absent octopamine. 

\item Series of 0.2 sec odor stimulations delivered coincident with sugar reward (which delivers octopamine). 
This gave data re how PN odor response is modulated by octopamine, relative to octopamine-free spontaneous FR.
See Figure \ref{pnTimecoursesJeff} panel A.

\item Series of 0.2 sec odor stimulations, delivered first without and then coincident with an octopamine wash applied to the AL.
This gave data re how PN spontaneous FR and PN odor response are modulated by octopamine.
See Figure \ref{pnTimecoursesJeff} panel B.

\end{enumerate}
In most cases the applied odor consisted of a collection of 5 volatiles, which taken together stimulate many glomeruli in the AL.
It was selected to ensure sufficient odor-responsive PNs, such that inserted electrodes would detect interesting (ie responsive) PNs.
Further details re \textit{in vivo} data collection can be found in \cite{schlizerman2014} and in Sec.~\ref{mothDataList}.
Example timecourses are shown in Figure \ref{pnTimecoursesJeff}.
 
\subsection{Simulation setup}~\\ \label{simulationSequence}
For Network Model learning experiments, the time sequence of events for simulations, shown in Fig \ref{schematicPlusTimecourses}, is as follows: 
\be
\item A period of no stimulus, to assess baseline spontaneous behavior. 
\item Four odor stimuli are delivered, 16 stimulations each (two odors were used in MB sparseness experiments). 
\item A period of control octopamine, ie without odor or Hebbian training.
\item The system is trained (odor + octopamine + Hebbian mechanism) on one of the odors. 
\item A period of no stimulus, to assess post-training spontaneous behavior. 
\item  The odors are re-applied (16 stimulations each), without octopamine, to assess effects of training on odor response.
\ee

%----------------------------------------------------

\clearpage

%%%%%%%%%%%%%%%%%%%%%%%%%%%%%%%%%%%%%%%%%%%%%
%\section*{Acknowledgements}

%%%%%%%%%%%%%%%%%%%%%%%%%%%%%%%%%%%%%%%%%%%%%
%\section*{Author Contributions}

%\newpage

%%%%%%%%%%%%%%%%%%%%%%%%%%%%%%%%%%

%%%%%%%%%%%%%%%%%%%%%%%%%%%%%%%%%%%%%%%%%%%%%%%%%%%%%%%%%%%%%%%

%																SUPPORTING INFORMATION

\section{Supporting Information}~\\
% \paragraph*{S1}
\subsection{Moth AL neural recording datasets}~\\ \label{mothDataList}
\paragraph*{Multichannel Recording Methods}
Adult male \textit{M. sexta} were reared in the laboratory at the University of Washington on an artificial diet under a long-day (17/7 hr light/dark cycle) photoperiod and prepared for experiments 2–3 d after emergence. In preparation for recording, the moth was secured in a plastic tube with dental wax, and the caudal end of the head capsule was opened and the piece of cuticle that is the attachment site of the pharyngeal dilator muscles is positioned forward and readhered to the headcapsule with myristic acid. The proboscis was extended and adhered to a piece of teflon tubing, 7-cm in length, that allowed movement and extension of the proboscis along its length and at the tip. Behavioral response to odor stimuli was determined by the percentage of proboscis movement response between treatments, allowing for a conservative estimate of a learned response. This preparation allows full access to the exposed AL while having no effect on the moth’s ability to feed normally. During recordings, the brain was superfused slowly with physiological saline solution (in mM: 150 NaCl, 3 CaCl2, 3 KCl, 10 N-tris[hydroxymethyl] methyl-2 aminoethanesulfonic acid buffer, and 25 sucrose, pH 6.9) throughout the experiment. 

Olfactory stimuli were delivered to the preparation by pulses of air from a constant air stream were diverted through a glass syringe containing a piece of filter paper bearing odor stimuli. The stimulus was pulsed by means of a solenoid-activated valve controlled by the SciWorks acquisition software (Data Wave Technologies, Longmont, CO, USA). The outlet of the stimulus syringe was positioned 2 cm from and orthogonal to the center of the antennal flagellum ipsilateral to the ALs. 

We used a classic conditioning paradigm to examine the effects on AL neurons while the moth learned to associate an odor with a sugar reward.  Odor stimuli were delivered in a three second pulses, and one second after odor onset, the unconditioned stimulus (US; 1 μL of 20\% sucrose solution) was applied to the proboscis for ca. 2 s. A ten minute inter-trial interval separated each training trial, and moths were trained over eight trials. After the conditioning trials were completed, a test trial was performed during which only the trained odor was presented to assess the behavioral and odor-driven ensemble responses as a result of the conditioning treatment. 

AL Recordings were made with 16-channel silicon multielectrode recording arrays (a4×4-3mm50-177; NeuroNexus Technologies, Ann Arbor, MI, USA). The spatial distribution design of the recording array suits the dimensions of the AL in \textit{M. sexta}, with four shanks spaced 125 μm apart, and each with four recording sites 50 μm apart. The four shanks were oriented in a line parallel to the antennal nerve. The probe was advanced slowly through the AL using a micromanipulator (Leica Microsystems, Bannockburn, IL, USA) until the uppermost recording sites were just below the surface of the AL. Extracellular activity was acquired with a RZ2 base station (Tucker-Davis Technologies, Alachua, FL, USA) and a RP2.1 real time processor (Tucker-Davis Technologies) and extracellular activity in the form of action potentials, or spikes, were extracted from the recorded signals and digitized at 25 kHz using the Tucker-Davis Technologies data-acquisition software.% (14, 30). 
Threshold and gain settings were adjusted independently for each channel, and spikes were captured in the 4-channel, or `tetrode’, recording configuration: any spike that passed threshold on one channel triggered the capture of spikes recorded on the other 3 channels on the same shank. Offline Sorter v.3 (Plexon Neurotechnology Research Systems, Dallas, TX, USA) was used to sort extracellular spikes based on their waveform shape and spikes were assigned timestamps to create raster plots and calculate peri-stimulus time histograms (PSTH). Only those clusters that were separated in three dimensional space (PC1–PC3) after statistical verification (multivariate ANOVA; P $<$ 0.05) were used for further analysis (typically 6–18 units were isolated per ensemble). Each spike in each cluster was time-stamped, and these data were used to create raster plots and to calculate peristimulus time histograms (PSTHs), interspike interval histograms, cross-correlograms, and rate histograms.\newline 

\paragraph*{List of \textit{in vivo} data sets}
\be
\item AL, odor only: PNs, one odor, no octopamine. 7 preps with 8 - 16 PNs each.
\item AL, odor  + octopamine. PNs, one odor, sugar reward. 10 preps with 9 - 21 PNs each.
\item AL + MB, odor + octopamine. PNs and KCs, one odor, sugar reward. 1 prep, with 7 PNs and 12 KCs.
\item AL, odor + octo wash: PNs, one odor, octopamine directly applied to AL. 7 preps: 6 preps with 8 - 13 PNs each; 1 prep with one pheromone-responsive neuron
\item AL, odors only (BEA): PNs, several odors and concentrations. 12 preps with 14 - 17 PNs each.
\item AL, odors only (ESO): PNs, several odors and concentrations. 4 preps with 12 - 14 PNs each.\newline
\ee

%\paragraph*{S2}
\subsection{Full equations of model dynamics}~\\ \label{fullDynamics}
\begin{eqnarray}
\tau_{\text{{\tiny R}}} \cdot d \bold{u}^{\text{{\tiny R}}}  &=&   f_{\text{{\tiny R}}} \big( \bold{u}^{\text{{\tiny R}}} , \bold{u}^{\text{{\tiny L}}} , \bold{u}^{\text{{\tiny S}}} ,  M^{\text{{\tiny L,R}}}, M^{\text{{\tiny S,R}}} , M^{\text{{\tiny O,R}}},o(t) \big) ~+~ d \bold{W}^{\text{{\tiny R}}} \\
\vspace{.3em}
\tau_{\text{{\tiny P}}} \cdot d \bold{u}^{\text{{\tiny P}}}  &=&   f_{\text{{\tiny P}}} \big( \bold{u}^{\text{{\tiny R}}} , \bold{u}^{\text{{\tiny P}}} , \bold{u}^{\text{{\tiny L}}}, M^{\text{{\tiny L,P}}}, M^{\text{{\tiny R,P}}} ,  M^{\text{{\tiny O,P}}},o(t) \big)  ~+~ d \bold{W}^{\text{{\tiny P}}} \\
\vspace{.3em}
\tau_{\text{{\tiny Q}}} \cdot d \bold{u}^{\text{{\tiny Q}}} &=&  f_{\text{{\tiny Q}}} \big( \bold{u}^{\text{{\tiny R}}} , \bold{u}^{\text{{\tiny Q}}} , \bold{u}^{\text{{\tiny L}}} , M^{\text{{\tiny L,Q}}}, M^{\text{{\tiny R,Q}}}, M^{\text{{\tiny O,Q}}},o(t) \big)  ~+~ d \bold{W}^{\text{{\tiny Q}}} \\
\vspace{.3em}
\tau_{\text{{\tiny L}}} \cdot d \bold{u}^{\text{{\tiny L}}} &=&  f_{\text{{\tiny L}}}\big ( \bold{u}^{\text{{\tiny R}}} , \bold{u}^{\text{{\tiny L}}},   M^{\text{{\tiny L,L}}}, M^{\text{{\tiny R,L}}}, M^{\text{{\tiny O,L}}},o(t) \big)  ~+~ d \bold{W}^{\text{{\tiny L}}} \\
\vspace{.3em}
\tau_{\text{{\tiny K}}} \cdot d \bold{u}^{\text{{\tiny K}}}  &=&  f_{\text{{\tiny K}}}\big ( \bold{u}^{\text{{\tiny P}}} , \bold{u}^{\text{{\tiny Q}}},    \bold{u}^{\text{{\tiny D}}}, M^{\text{{\tiny P,K}}}, M^{\text{{\tiny Q,K}}} \big) ~+~ d \bold{W}^{\text{{\tiny K}}} \\ 
\vspace{.3em}
\tau_{\text{{\tiny E}}} \cdot d \bold{u}^{\text{{\tiny E}}} &=&  f_{\text{{\tiny E}}}\big ( \bold{u}^{\text{{\tiny K}}} , \bold{u}^{\text{{\tiny E}}},  M^{\text{{\tiny K,E}}} \big) 
\end{eqnarray}
where
\begin{equation*}
\begin{cases}
f_{\text{{\tiny R}}}  =  - \bold{u}^{\text{{\tiny R}}}  + \text{sigmoid} \big [ -\big (I- \gamma \cdot o(t) \cdot M^{\text{{\tiny O,R}}} \big ) 
M^{\text{{\tiny L,R}}} \hspace{.2em} {\bf{u}}^{\text{{\tiny L}}} ~+~ \big ( I + o(t)\cdot M^{\text{{\tiny O,R}}} \big) M^{\text{{\tiny S,R}}} \hspace{.2em}  \bf{u}^{\text{{\tiny S}}}
\big ]  \\
\vspace{.3em}
f_{\text{{\tiny P}}}  =  - \bold{u}^{\text{{\tiny P}}}  + \text{sigmoid} \big [ -\big (I- \gamma \cdot o(t) \cdot M^{\text{{\tiny O,P}}} \big )  M^{\text{{\tiny L,P}}}  \hspace{.2em}  {\bf{u}}^{\text{{\tiny L}}} 
~+~ \big ( I + o(t)\cdot M^{\text{{\tiny O,P}}} \big) M^{\text{{\tiny R,P}}} \hspace{.2em}  \bf{u}^{\text{{\tiny R}}} \big ]  \\
\vspace{.3em}
f_{\text{{\tiny Q}}}  =  - \bold{u}^{\text{{\tiny Q}}}  + \text{sigmoid} \big [ -\big (I- \gamma \cdot o(t) \cdot M^{\text{{\tiny O,Q}}} \big )  M^{\text{{\tiny L,Q}}}  \hspace{.2em}  {\bf{u}}^{\text{{\tiny L}}} 
~+~ \big ( I + o(t)\cdot M^{\text{{\tiny O,Q}}} \big) M^{\text{{\tiny R,Q}}} \hspace{.2em}  \bf{u}^{\text{{\tiny R}}} \big ] \\
\vspace{.3em}
f_{\text{{\tiny L}}}  =  - \bold{u}^{\text{{\tiny L}}}  + \text{sigmoid} \big [ -\big (I- \gamma \cdot o(t) \cdot M^{\text{{\tiny O,L}}} \big ) 
M^{\text{{\tiny L,L}}} \hspace{.2em}  {\bf{u}}^{\text{{\tiny L}}} ~+~ \big ( I + o(t)\cdot M^{\text{{\tiny O,L}}} \big) M^{\text{{\tiny R,L}}} \hspace{.2em}  \bf{u}^{\text{{\tiny R}}}
\big ]  \\
\vspace{.3em}
f_{\text{{\tiny K}}}  =  - \bold{u}^{\text{{\tiny K}}}  + \text{sigmoid} \big [ -\big ( {\bf{u}}^{\text{{\tiny D}}} + M^{\text{{\tiny Q,K}}} \hspace{.2em} {\bf{u}}^{\text{{\tiny Q}}} \big)  ~+~   M^{\text{{\tiny P,K}}} \hspace{.2em} \bf{u}^{\text{{\tiny P}}}
\big ]  \\
\vspace{.3em}
f_{\text{{\tiny E}}}  =  - \bold{u}^{\text{{\tiny E}}}  + M^{\text{{\tiny K,E}}} \hspace{.2em}  \bf{u}^{\text{{\tiny K}}}   \\
\end{cases}
\end{equation*}

\begin{table}
\centering
\caption{Variables and parameters for neuronal network model}
\vspace{1em}
%\footnotesize
\scriptsize
\begin{tabular}{ c c c l }
%\hline
Symbol & Type & Size/Value & Description and Remarks
 \\ \hline
 \\
R & superscript & & Refers to the \textit{receptor neurons} subpopulation. \\
P & superscript & & Refers to the \textit{excitatory projection neurons} subpopulation. \\
Q & superscript & & Refers to the \textit{inhibitory projection neurons} subpopulation.  \\
L & superscript & & Refers to the \textit{lateral neurons} subpopulation. \\
K & superscript & & Refers to the \textit{kenyon cells} subpopulation.  \\
E & superscript & & Refers to the readout \textit{extrinsic neurons} subpopulation. \\
O & superscript & & Refers to the \textit{octopamine} neurotransmitter. \\
\\
$nG$ & scalar & 60 & Number of glomeruli in the antenna lobe. $^{*}$ \\
$nS$ & scalar & 2-4 & Number of different stimuli (odors).  \\
$nQ$ & scalar &  & Number of inhibitory projection neurons. \\
$nK$ & scalar & 2000 & Number of kenyon cells. \\
$nE$ & scalar & 1 & Number of extrinsic neurons. \\

\\
 $\bold{u}^{\text{{\tiny R}}}$ & vector &  $nG \times 1$ & FRs of the receptor neurons subpopulation. \\
 $\bold{u}^{\text{{\tiny P}}}$ & vector &  $nG \times 1$ & FRs of the exc. projection neurons subpopulation.\\
 $\bold{u}^{\text{{\tiny Q}}}$ & vector &  $nQ \times 1$ & FRs of the inh. projection neurons subpopulation. \\
 $\bold{u}^{\text{{\tiny L}}}$ & vector &  $nG \times 1$ & FRs of the lateral neurons subpopulation. \\
 $\bold{u}^{\text{{\tiny K}}}$ & vector &  $nK \times 1$ & FRs of the kenyon cells subpopulation. Sparse. \\
 $\bold{u}^{\text{{\tiny E}}}$ & vector &  $nE \times 1$ & FRs of the extrinsic neurons subpopulation. \\
 \\
$\bold{u}^{\text{{\tiny S}}}$ & vector &  &  \\
$\bold{u}^{\text{{\tiny D}}}$ & vector &  &  \\
\\
$M^{\text{{\tiny S,R}}}$ & matrix & $nG \times nS$ & Stimulus $\rightarrow \bold{u}^{\text{{\tiny R}}}$ connections.  \\
\\
$M^{\text{{\tiny O,R}}}$ & matrix & $nG \times nG$ & Octopamine $\rightarrow \bold{u}^{\text{{\tiny R}}}$ connections. Diagonal matrix.\\
$M^{\text{{\tiny O,L}}}$ & matrix & $nG \times nG$ & Octopamine $\rightarrow \bold{u}^{\text{{\tiny L}}}$ connections. Diagonal matrix. \\
\\
$M^{\text{{\tiny R,L}}}$ & matrix & $nG \times nG$ & Connection weights ~ $\bold{u}^{\text{{\tiny R}}} \rightarrow \bold{u}^{\text{{\tiny L}}}$.\\
$M^{\text{{\tiny R,P}}}$ & matrix & $nG \times nG$ & Connection weights ~ $\bold{u}^{\text{{\tiny R}}} \rightarrow \bold{u}^{\text{{\tiny P}}}$. Diagonal matrix.\\
$M^{\text{{\tiny R,Q}}}$ & matrix & $nQ \times nG$ & Connection weights ~ $\bold{u}^{\text{{\tiny R}}} \rightarrow \bold{u}^{\text{{\tiny Q}}}$.\\
$M^{\text{{\tiny P,K}}}$ & matrix & $nK \times nG$ & Connection weights ~ $\bold{u}^{\text{{\tiny P}}} \rightarrow \bold{u}^{\text{{\tiny K}}}$.\\
$M^{\text{{\tiny Q,K}}}$ & matrix & $nK \times nQ$ & Connection weights ~ $\bold{u}^{\text{{\tiny Q}}} \rightarrow \bold{u}^{\text{{\tiny K}}}$.\\
$M^{\text{{\tiny L,R}}}$ & matrix & $nG \times nG$ & Connection weights ~ $\bold{u}^{\text{{\tiny L}}} \rightarrow \bold{u}^{\text{{\tiny R}}}$. \\
$M^{\text{{\tiny L,P}}}$ & matrix & $nG \times nG$ & Connection weights ~ $\bold{u}^{\text{{\tiny L}}} \rightarrow \bold{u}^{\text{{\tiny P}}}$.\\
$M^{\text{{\tiny L,Q}}}$ & matrix & $nQ \times nG$& Connection weights ~ $\bold{u}^{\text{{\tiny L}}} \rightarrow \bold{u}^{\text{{\tiny Q}}}$. \\
$M^{\text{{\tiny L,L}}}$ & matrix & $nG \times nG$ & Connection weights ~ $\bold{u}^{\text{{\tiny L}}} \rightarrow \bold{u}^{\text{{\tiny L}}}$. \\
$M^{\text{{\tiny K,E}}}$ & matrix & $nE \times nK$ & Connection weights ~ $\bold{u}^{\text{{\tiny K}}} \rightarrow \bold{u}^{\text{{\tiny E}}}$. \\
\\
$o(t)$ & function &  0 or 1 & Flags when octopamine is active (typically during training).\\
$\gamma $ & scalar & 0.5 & Scaling factor for octopamine's effects on inhibition. $^{*}$ \\
\\
$\tau_{\text{{\tiny R}}}$ & scalar & & \\
$\tau_{\text{{\tiny P}}}$ & scalar & & \\
$\tau_{\text{{\tiny Q}}}$ & scalar & & \\
$\tau_{\text{{\tiny L}}}$ & scalar & & \\
$\tau_{\text{{\tiny K}}}$ & scalar & & \\
$\tau_{\text{{\tiny E}}}$ & scalar & & \\\\
\hline
\end{tabular}
\\
{\tiny $^{*}$  Each glomerulus receives one RN and one octopamine input, and initiates one PN and one LN. \\
$^{*}$ Octopamine decreases the response to inhibition less than it increases the response to excitation}
\label{NetSymbols}
\end{table}

\clearpage

% \paragraph*{S3}
% \subsection*{Learning with unequal magnitude odors} \label{unequalMag4OdorLearningFig}

%\clearpage

\subsection{ANOVA analysis of Network Model learning}
\begin{figure}[h!t] 
	%\begin{adjustwidth} {-60mm}{0mm}
	\centering
	\includegraphics[width=120mm] {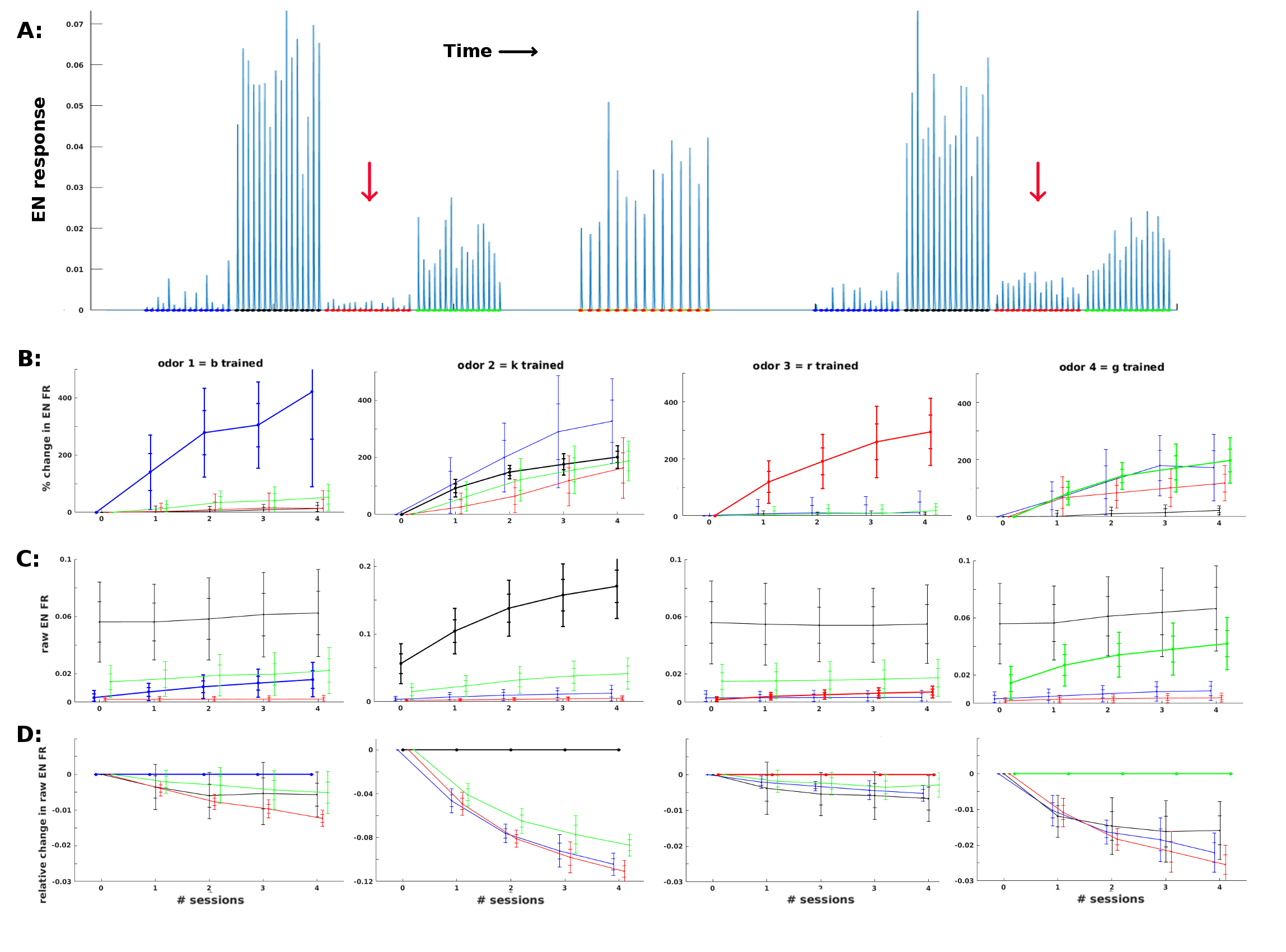}    
	\caption { {\bf Effect of training on EN FRs, given odors with unequal naive response magnitudes.}
		When odors induced naive EN responses of very different magnitudes, then trained odor response increased much more than control odor responses either in raw magnitude, or as a percentage, or both.\newline
		\textbf{A:} Typical timecourse showing magnitudes EN responses before and after training the third (red) odor, indicated by red arrow, over 15 odor stimulations. 
		This corresponds to the third column in panels B - D, at index 3 on the x-axis.
		Note that only the third (red) odor's EN response changes magnitude. \newline
		Panels B - D: Changes to ENs during training. x-axis = number of training sessions. 
		Each column shows results of training a given odor, color coded: blue, black, red. 
		y-axis measures raw EN or percent change in EN. 
		21 trials per data point. \newline
		\textbf{B:}  Percent change (from pre-training)  in ENs, mean $\pm$2 stds. \newline
		\textbf{C:} Raw EN FRs, mean $\pm$2 stds.\newline
		\textbf{D:} Changes in raw EN FRs, normalized by trained odor (ie subtract the trained odor's changes from all odors), mean $\pm$2 std devs. This shows how far each control odor lagged behind the trained odor.\newline
		Note that the trained odor dominates in either raw increase (panels C, D) if naive response to trained odor was large, or in percent increase (panel B) if naive response to trained odor was small.
	} 
	\label{unequalMag4OdorResults} 
	%\end{adjustwidth}
\end{figure}

~\\
\label{anova}
% {\textbf{ANOVA analysis of Network Model learning}}
The differential increase in EN response to trained vs control odors was almost always significant to $p  < 0.01$.
When odors' naive EN response magnitudes differed by $>$ 3, sometimes differences in either raw increases or percentage increases (not both) did not attain this level of significance, while the other metric did.
An example of learning results when odors' naive EN response magnitudes differ widely is shown in Fig \ref{unequalMag4OdorResults}.

\begin{figure}[t]

%\begin{adjustwidth} {-50mm}{0mm}
\centering
\includegraphics [width=120mm] {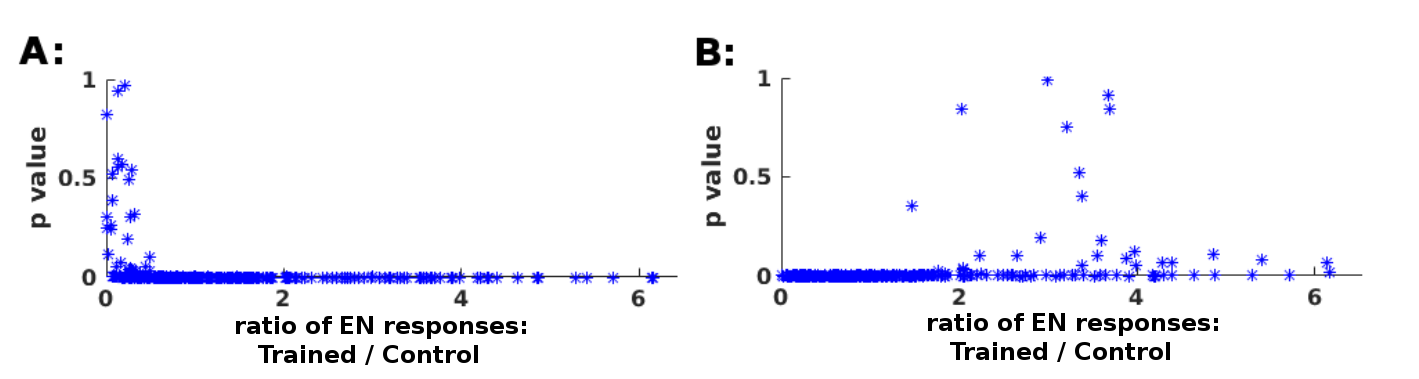}  
\caption{ {\bf p-values for trained-control odor pairs:} % 
{\bf{A:}} p-values for change in raw EN responses. 
{\bf{B:}} p-values for percentage change in EN responses.
P-values are sometimes high (for one metric or the other) when trained and control odors have highly disparately-scaled naive responses $\mu_T$ (= mean raw T ) $\mu_C$ (= mean raw C). 
Plots show results given 20 training stimulations.\newline
When $\mu_T$ is larger (right end of x-axis), the p-value for raw change (A) is consistently very low, but the p-value for percentage change (rB) can be high, since even a small incidental change to a low-intensity odor can be a large percentage change.\newline
When $\mu_C$ is larger (left end of x-axis), the p-value for percentage change (B) is consistently very low, but the p-value for raw change (A) can be high, since even a small percentage change to a high-response odor corresponds to a large raw change.\newline
When naive odor responses are roughly matched, eg within 3x (ie 0.33 to 3), p-values for both raw and percentage change are very low.
 }
\label{anovaScatterplot}
%\end{adjustwidth} 
\end{figure}

The p-values of 336 trained odor/control odor pairs are plotted in Fig \ref{anovaScatterplot}, against the ratio of their mean naive odor responses $\frac{\mu_T}{\mu_C}$, for 28 moths randomly generated from a template, with three control odors and one trained odor.
Each p-value is for the trained odor vs one control odor (so there are 12 data points per moth). 
Column 1 shows p-values for the change in raw EN response, as in Fig \ref{unequalMag4OdorResults} (C), trained vs control. 
Trained odors with very low-magnitude naive response often did not have raw increases larger than high-magnitude control odors.
Column 2  shows p-values for the percentage change in EN response, as in Fig \ref{unequalMag4OdorResults} (B), trained vs control.

\begin{figure}[t]

%  \begin{adjustwidth} {-50mm}{0mm}
\centering
\includegraphics [width=120mm] {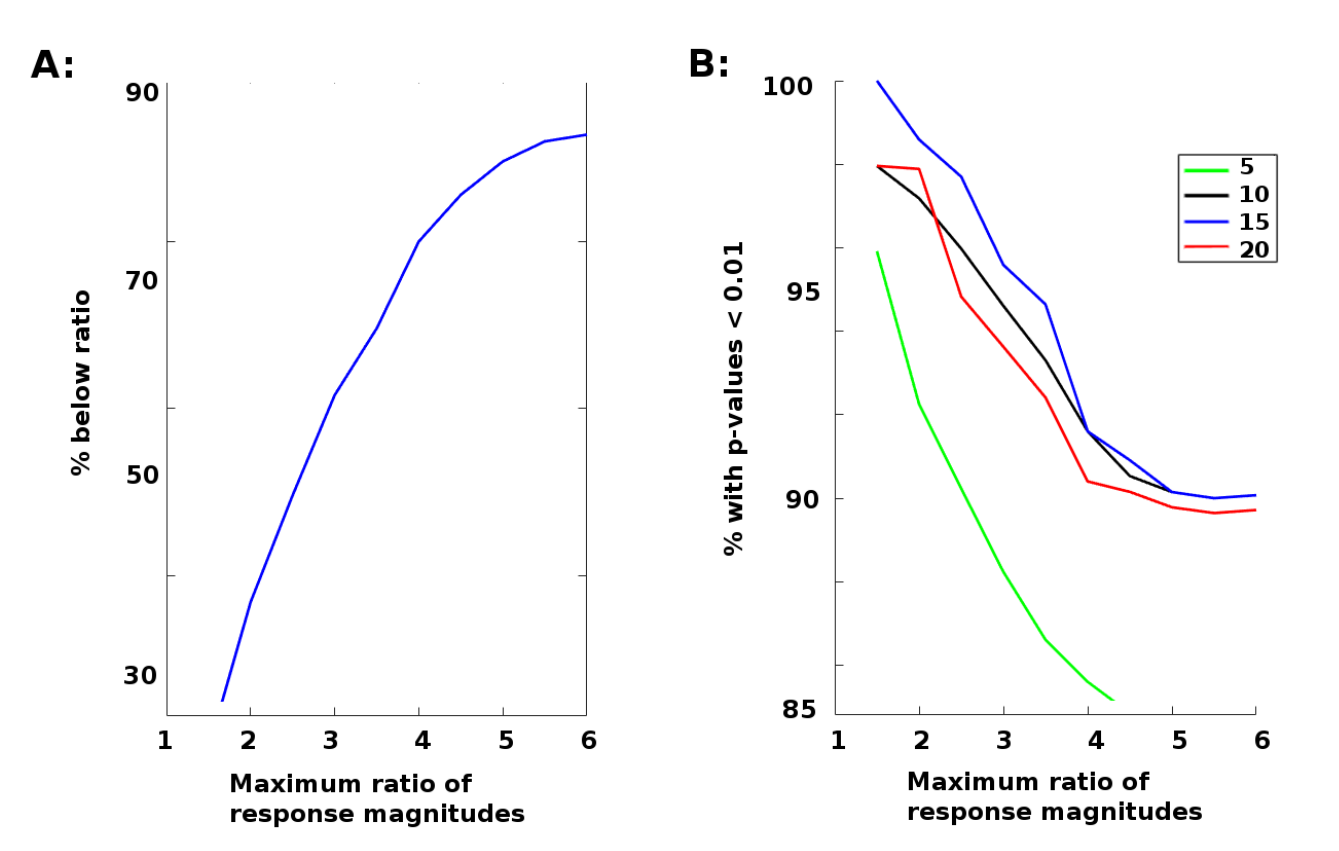} 
\caption{  {\bf Fractions of p-values below 0.01 for trained-control odor pairs}
In most cases, the trained odor shows much larger increases in EN response magnitude.
{\bf{A:}} The percentage of trained-control odor pairs with EN response magnitudes within the ratios given on the x-axis.
{\bf{B:}} The percentage of trained-control pairs, with EN response magnitudes within the ratios given on the x-axis, whose training-induced changes in EN responses were distinct with p-value $<$ 0.01.
Each curve is for a different number of training stimulations.
More training increases distinctions, up to 15 stimulations. But additional training actually hinders distinctions, as control odor response reinforcement begins to overtake trained odor reinforcement.
 }
\label{anovaLinePlots}
 %\end{adjustwidth}
\end{figure}

Unless the naive EN responses for the two odors were highly disparate (eg by factor of $>$3x), the differential increase in EN response of the trained vs control odors is almost always significant, measured both as raw and as percentage.
Fig \ref{anovaLinePlots} plots the percentage of 336 trained-control pairs that had p-values for both measures of EN response increase (ie as raw and as percentage) below the listed threshold (eg $p = 0.01$),  for 336 trained-control pairs whose ratio ($\frac{\mu_T}{\mu_C}$ or $\frac{\mu_C}{\mu_T}$) is within the bound given on the x-axis. 
Fig \ref{anovaLinePlots} shows how many moths, generated from template with no constraint on unbalanced naive odor EN responses, had differential post-training EN responses with significance $p < 0.01$, for both measures (as raw and as a percentage).

%------------------------------------------

\clearpage

%\nolinenumbers 

%\bibliographystyle{plos2015.bst}
\bibliographystyle{unsrt}
\bibliography{mothBibliography_nov2017_includesPedroBib}

\begin{thebibliography}{10}

\bibitem{hammerMenzel1998}
Martin Hammer and Randolf Menzel.
\newblock Multiple sites of associative odor learning as revealed by local
  brain microinjections of octopamine in honeybees.
\newblock {\em Learn Mem}, 5(1):146--156, May 1998.
\newblock 10454379[pmid].

\bibitem{Riffell2008}
Jeffrey~A. Riffell, Leif Abrell, and John~G. Hildebrand.
\newblock Physical processes and real-time chemical measurement of the insect
  olfactory environment.
\newblock {\em Journal of Chemical Ecology}, 34(7):837--853, Jul 2008.

\bibitem{wilson2008}
Rachel~I Wilson.
\newblock Neural and behavioral mechanisms of olfactory perception.
\newblock {\em Current Opinion in Neurobiology}, 18(4):408 -- 412, 2008.
\newblock Sensory systems.

\bibitem{campbellMushroomBody}
Robert~A.A. Campbell and Glenn~C. Turner.
\newblock The mushroom body.
\newblock {\em Current Biology}, 20(1):R11 -- R12, 2010.

\bibitem{dacksRiffell2009}
{Andrew M.} Dacks, {Jeffrey A.} Riffell, {Joshua P.} Martin, {Stephanie L.}
  Gage, and {Alan J.} Nighorn.
\newblock Olfactory modulation by dopamine in the context of aversive learning.
\newblock {\em Journal of Neurophysiology}, 108(2):539--550, 7 2012.

\bibitem{masse2009}
Nicolas~Y. Masse, Glenn~C. Turner, and Gregory~S.X.E. Jefferis.
\newblock Olfactory information processing in drosophila.
\newblock {\em Current Biology}, 19(16):R700 -- R713, 2009.

\bibitem{laurent2002}
Gilles Laurent.
\newblock Olfactory network dynamics and the coding of multidimensional
  signals.
\newblock {\em Nature Reviews Neuroscience}, 3:884 EP --, Nov 2002.
\newblock Review Article.

\bibitem{honeggerTurner2011}
Kyle~S. Honegger, Robert A.~A. Campbell, and Glenn~C. Turner.
\newblock Cellular-resolution population imaging reveals robust sparse coding
  in the drosophila mushroom body.
\newblock {\em Journal of Neuroscience}, 31(33):11772--11785, 2011.

\bibitem{caron2016}
Sophie J.~C. Caron.
\newblock Brains don{\textquoteright}t play dice{\textemdash}or do they?
\newblock {\em Science}, 342(6158):574--574, 2013.

\bibitem{galizia2014}
C.~Giovanni Galizia.
\newblock Olfactory coding in the insect brain: data and conjectures.
\newblock {\em European Journal of Neuroscience}, 39(11):1784--1795, 2014.

\bibitem{peng2017}
Fei Peng and Lars Chittka.
\newblock A simple computational model of the bee mushroom body can explain
  seemingly complex forms of olfactory learning and memory.
\newblock {\em Current Biology}, 27(2):224 -- 230, 2017.

\bibitem{roper2017}
Mark Roper, Chrisantha Fernando, and Lars Chittka.
\newblock Insect bio-inspired neural network provides new evidence on how
  simple feature detectors can enable complex visual generalization and
  stimulus location invariance in the miniature brain of honeybees.
\newblock {\em PLOS Computational Biology}, 13(2):1--23, 02 2017.

\bibitem{mosqueiro2014}
Thiago~S Mosqueiro and Ramón Huerta.
\newblock Computational models to understand decision making and pattern
  recognition in the insect brain.
\newblock {\em Current Opinion in Insect Science}, 6:80 -- 85, 2014.
\newblock Pests and resistance/Parasites/Parasitoids/Biological
  control/Neurosciences.

\bibitem{arena2013}
Paolo Arena, Luca Patané, Vincenzo Stornanti, Pietro~Savio Termini, Bianca
  Zäpf, and Roland Strauss.
\newblock Modeling the insect mushroom bodies: Application to a delayed
  match-to-sample task.
\newblock {\em Neural Networks}, 41:202 -- 211, 2013.
\newblock Special Issue on Autonomous Learning.

\bibitem{faghihi2017}
Faramarz Faghihi, Ahmed~A. Moustafa, Ralf Heinrich, and Florentin Wörgötter.
\newblock A computational model of conditioning inspired by drosophila
  olfactory system.
\newblock {\em Neural Networks}, 87:96 -- 108, 2017.

\bibitem{dayan2001}
Peter Dayan and L.~F. Abbott.
\newblock {\em Theoretical Neuroscience: Computational and Mathematical
  Modeling of Neural Systems}.
\newblock The MIT Press, 2005.

\bibitem{hubel}
D.~Hubel and T.~Wiesel.
\newblock Receptive fields, binocular interaction, and functional architecture
  in the cat's visual cortex.
\newblock {\em Journal of Physiology}, 160:106--154, 1962.

\bibitem{fukushima}
Kunihiko Fukushima.
\newblock {N}eocognitron: {A} self-organizing neural network model for a
  mechanism of pattern recognition unaffected by shift in position.
\newblock {\em Biological Cybernetics}, 36:193--202, 1980.

\bibitem{schmidhuber2014}
Jurgen Schmidhuber.
\newblock Deep learning in neural networks: An overview.
\newblock {\em Neural Networks}, 61(Supplement C):85 -- 117, 2015.

\bibitem{lecunIeeeSpectrum}
Yann LeCun.
\newblock Facebook ai director yann lecun on his quest to unleash deep learning
  and make machines smarter.
\newblock {\em IEEE Spectrum}, 2015.

\bibitem{martin2011}
Joshua~P. Martin, Aaron Beyerlein, Andrew~M. Dacks, Carolina~E. Reisenman,
  Jeffrey~A. Riffell, Hong Lei, and John~G. Hildebrand.
\newblock The neurobiology of insect olfaction: Sensory processing in a
  comparative context.
\newblock {\em Progress in Neurobiology}, 95(3):427 -- 447, 2011.

\bibitem{kvello2009}
Pal Kvello, Bjarte Lofaldli, Jurgen Rybak, Randolf Menzel, and Hanna
  Mustaparta.
\newblock Digital, three-dimensional average shaped atlas of the heliothis
  virescens brain with integrated gustatory and olfactory neurons.
\newblock {\em Frontiers in Systems Neuroscience}, 3:14, 2009.

\bibitem{bhandawat2007}
Vikas Bhandawat, Shawn~R Olsen, Nathan~W Gouwens, Michelle~L Schlief, and
  Rachel~I Wilson.
\newblock Sensory processing in the drosophila antennal lobe increases
  reliability and separability of ensemble odor representations.
\newblock {\em Nature Neuroscience}, 10:1474--1482, 2007.

\bibitem{perisse2013}
Emmanuel Perisse, Christopher Burke, Wolf Huetteroth, and Scott Waddell.
\newblock Shocking revelations and saccharin sweetness in the study of
  drosophila olfactory memory.
\newblock {\em Curr Biol}, 23(17):R752--R763, Sep 2013.
\newblock S0960-9822(13)00921-4[PII], 24028959[pmid].

\bibitem{bazhenovStopfer2010}
Maxim Bazhenov and Mark Stopfer.
\newblock Forward and back: Motifs of inhibition in olfactory processing.
\newblock {\em Neuron}, 67(3):357 -- 358, 2010.

\bibitem{campbell2013}
RAA Campbell, KS~Honegger, H~Qin, W~Li, E~Demir, and GC~Turner.
\newblock Imaging a population code for odor identity in the drosophila
  mushroom body.
\newblock {\em Journal of Neuroscience}, 33(25):10568--81, 2013.

\bibitem{hige2015}
Toshihide Hige, Yoshinori Aso, Gerald~M. Rubin, and Glenn~C. Turner.
\newblock Plasticity-driven individualization of olfactory coding in mushroom
  body output neurons.
\newblock {\em Nature}, 526:258 EP --, Sep 2015.

\bibitem{hammer1995}
M~Hammer and R~Menzel.
\newblock Learning and memory in the honeybee.
\newblock {\em Journal of Neuroscience}, 15(3):1617--1630, 1995.

\bibitem{cassenaer}
Stijn Cassenaer and Gilles Laurent.
\newblock Hebbian stdp in mushroom bodies facilitates the synchronous flow of
  olfactory information in locusts.
\newblock {\em Nature}, 448:709 EP --, Jun 2007.

\bibitem{turner2008}
Glenn~C. Turner, Maxim Bazhenov, and Gilles Laurent.
\newblock Olfactory representations by drosophila mushroom body neurons.
\newblock {\em Journal of Neurophysiology}, 99(2):734--746, 2008.

\bibitem{ganguli2012}
Surya Ganguli and Haim Sompolinsky.
\newblock Compressed sensing, sparsity, and dimensionality in neuronal
  information processing and data analysis.
\newblock {\em Annual Review of Neuroscience}, 35(1):485--508, 2012.
\newblock PMID: 22483042.

\bibitem{litwinKumarHarris2017}
Ashok Litwin-Kumar, Kameron~Decker Harris, Richard Axel, Haim Sompolinsky, and
  L.F. Abbott.
\newblock Optimal degrees of synaptic connectivity.
\newblock {\em Neuron}, 93(5):1153 -- 1164.e7, 2017.

\bibitem{olsenWilson2010}
Shawn~R. Olsen, Vikas Bhandawat, and Rachel~Irene Wilson.
\newblock Divisive normalization in olfactory population codes.
\newblock {\em Neuron}, 66(2):287--299, Apr 2010.
\newblock 20435004[pmid].

\bibitem{Lei2002}
Hong Lei, Thomas~A. Christensen, and John~G. Hildebrand.
\newblock Local inhibition modulates odor-evoked synchronization of
  glomerulus-specific output neurons.
\newblock {\em Nature Neuroscience}, 5:557 EP --, May 2002.
\newblock Article.

\bibitem{babadi}
Baktash Babadi and Haim Sompolinsky.
\newblock Sparseness and expansion in sensory representations.
\newblock {\em Neuron}, 83(5):1213 -- 1226, 2014.

\bibitem{suttonBarto}
Richard~S Sutton and Andrew~G Barto.
\newblock {\em Reinforcement Learning}.
\newblock MIT Press, 1998.

\bibitem{settles}
B.~Settles.
\newblock {\em Active Learning}.
\newblock Synthesis Lectures on Artificial Intelligence and Machine Learning.
  Morgan \& Claypool, 2012.

\bibitem{attenberg}
Josh Attenberg and Foster Provost.
\newblock Why label when you can search?: Alternatives to active learning for
  applying human resources to build classification models under extreme class
  imbalance.
\newblock In {\em Proceedings of the 16th ACM SIGKDD International Conference
  on Knowledge Discovery and Data Mining}, KDD '10, pages 423--432, New York,
  NY, USA, 2010. ACM.

\bibitem{maiaReactionTime}
Pedro~D. Maia and J.~Nathan Kutz.
\newblock Reaction time impairments in decision-making networks as a diagnostic
  marker for traumatic brain injuries and neurological diseases.
\newblock {\em Journal of Computational Neuroscience}, 42:323--347, 2017.

\bibitem{hong2015}
Elizabeth~J. Hong and Rachel~I. Wilson.
\newblock Simultaneous encoding of odors by channels with diverse sensitivity
  to inhibition.
\newblock {\em Neuron}, 85(3):573 -- 589, 2015.

\bibitem{maPougetBayesianInference}
Wei~Ji Ma, Jeffrey~M. Beck, Peter~E. Latham, and Alexandre Pouget.
\newblock Bayesian inference with probabilistic population codes.
\newblock {\em Nature Neuroscience}, 9:1432 EP --, Oct 2006.
\newblock Article.

\bibitem{guozhong}
Guozhong An.
\newblock The effects of adding noise during backpropagation training on a
  generalization performance.
\newblock {\em Neural Comput.}, 8(3):643--674, April 1996.

\bibitem{szegedy}
Christian Szegedy, Wojciech Zaremba, Ilya Sutskever, Joan Bruna, Dumitru Erhan,
  Ian~J. Goodfellow, and Rob Fergus.
\newblock Intriguing properties of neural networks.
\newblock {\em CoRR}, abs/1312.6199, 2013.

\bibitem{bengioNIPS2015}
Yoshua Bengio and Asja Fischer.
\newblock Early inference in energy-based models approximates back-propagation.
\newblock {\em arXiv e-prints}, abs/1510.02777, October 2015.

\bibitem{nagel2011}
Katherine~I. Nagel and Rachel~I. Wilson.
\newblock Biophysical mechanisms underlying olfactory receptor neuron dynamics.
\newblock {\em Nature Neuroscience}, 14:208 EP --, Jan 2011.
\newblock Article.

\bibitem{deisig2006}
Nina Deisig, Martin Giurfa, Harald Lachnit, and Jean-Christophe Sandoz.
\newblock Neural representation of olfactory mixtures in the honeybee antennal
  lobe.
\newblock {\em European Journal of Neuroscience}, 24(4):1161--1174, 2006.

\bibitem{riffellPnas2009}
Jeffrey~A. Riffell, H.~Lei, and John~G. Hildebrand.
\newblock Neural correlates of behavior in the moth manduca sexta in response
  to complex odors.
\newblock {\em Proceedings of the National Academy of Sciences},
  106(46):19219--19226, 2009.

\bibitem{riffellCurrBio2009}
Jeffrey~A. Riffell, Hong Lei, Thomas~A. Christensen, and John~G. Hildebrand.
\newblock Characterization and coding of behaviorally significant odor
  mixtures.
\newblock {\em Current Biology}, 19(4):335 -- 340, 2009.

\bibitem{wilson2005}
Rachel~I. Wilson and Gilles Laurent.
\newblock Role of gabaergic inhibition in shaping odor-evoked spatiotemporal
  patterns in the drosophila antennal lobe.
\newblock {\em Journal of Neuroscience}, 25(40):9069--9079, 2005.

\bibitem{olsen2008}
Shawn~R. Olsen, Vikas Bhandawat, and Rachel~I. Wilson.
\newblock Excitatory interactions between olfactory processing channels in the
  <em>drosophila</em> antennal lobe.
\newblock {\em Neuron}, 54(4):667, 2008.

\bibitem{olsenWilson2008}
Shawn~R. Olsen and Rachel~I. Wilson.
\newblock Lateral presynaptic inhibition mediates gain control in an olfactory
  circuit.
\newblock {\em Nature}, 452:956 EP --, Mar 2008.
\newblock Article.

\bibitem{sjoholm2006}
Marcus Sjoholm.
\newblock Structure and function of the moth mushroom body.
\newblock {\em Swedish Univ of Agricultural Sciences, Alrarp}, 2006.
\newblock PhD thesis.

\bibitem{jefferisPotter2007}
Gregory~S.X.E. Jefferis, Christopher~J. Potter, Alexander~M. Chan, Elizabeth~C.
  Marin, Torsten Rohlfing, Calvin R.~Maurer Jr., and Liqun Luo.
\newblock Comprehensive maps of drosophila higher olfactory centers: Spatially
  segregated fruit and pheromone representation.
\newblock {\em Cell}, 128(6):1187 -- 1203, 2007.

\bibitem{lin2014}
Andrew~C. Lin, Alexei~M. Bygrave, Alix de~Calignon, Tzumin Lee, and Gero
  Miesenb{\"o}ck.
\newblock Sparse, decorrelated odor coding in the mushroom body enhances
  learned odor discrimination.
\newblock {\em Nature Neuroscience}, 17:559 EP --, Feb 2014.
\newblock Article.

\bibitem{gruntman}
Eyal Gruntman and Glenn~C. Turner.
\newblock Integration of the olfactory code across dendritic claws of single
  mushroom body neurons.
\newblock {\em Nature Neuroscience}, 16:1821 EP --, Oct 2013.
\newblock Article.

\bibitem{luo2006}
Sean~X. Luo, Richard Axel, and L.~F. Abbott.
\newblock Generating sparse and selective third-order responses in the
  olfactory system of the fly.
\newblock {\em Proceedings of the National Academy of Sciences},
  107(23):10713--10718, 2010.

\bibitem{perezOrive}
Javier Perez-Orive, Ofer Mazor, Glenn~C. Turner, Stijn Cassenaer, Rachel~I.
  Wilson, and Gilles Laurent.
\newblock Oscillations and sparsening of odor representations in the mushroom
  body.
\newblock {\em Science}, 297(5580):359--365, 2002.

\bibitem{caron2013}
SJ~Caron, V~Ruta, LF~Abbott, and R~Axel.
\newblock Random convergence of olfactory inputs in the drosophila mushroom
  body.
\newblock {\em Nature}, 497(5):113--7, 2013.

\bibitem{menzelManz}
Randolf Menzel and Gisela Manz.
\newblock Neural plasticity of mushroom body-extrinsic neurons in the honeybee
  brain.
\newblock {\em Journal of Experimental Biology}, 208(22):4317--4332, 2005.

\bibitem{riffell2013}
Jeffrey~A. Riffell, Hong Lei, Leif Abrell, and John~G. Hildebrand.
\newblock Neural basis of a pollinator{\textquoteright}s buffet: Olfactory
  specialization and learning in manduca sexta.
\newblock {\em Science}, 2012.

\bibitem{silbering2007}
Ana~F. Silbering and C.~Giovanni Galizia.
\newblock Processing of odor mixtures in the drosophila antennal lobe reveals
  both global inhibition and glomerulus-specific interactions.
\newblock {\em Journal of Neuroscience}, 27(44):11966--11977, 2007.

\bibitem{higham2001}
Desmond~J. Higham.
\newblock An algorithmic introduction to numerical simulation of stochastic
  differential equations.
\newblock {\em SIAM Rev.}, 43(3):525--546, March 2001.

\bibitem{hebb}
D.~O. Hebb.
\newblock {\em The organization of behavior : a neuropsychological theory}.
\newblock Wiley New York, 1949.

\bibitem{grant}
W.~Shane Grant, James Tanner, and Laurent Itti.
\newblock Biologically plausible learning in neural networks with modulatory
  feedback.
\newblock {\em Neural Networks}, 88(Supplement C):32 -- 48, 2017.

\bibitem{barabasi1999}
Albert-Laszlo Barabasi and Reka Albert.
\newblock Emergence of scaling in random networks.
\newblock {\em Science}, 286(5439):509--512, 1999.

\bibitem{schlizerman2014}
Eli Shlizerman, Jeffrey~A. Riffell, and J.~Nathan Kutz.
\newblock Data-driven inference of network connectivity for modeling the
  dynamics of neural codes in the insect antennal lobe.
\newblock {\em Frontiers in Computational Neuroscience}, 8:70, 2014.

\end{thebibliography}

%%%%%%%%%%%%%%%%%%%%%%%%%%%%%%%%%%%%%%%%%%%%%%%%%%%%%%%%%%%%%%%%%
%%%%%%%%%%%%%%%%%%%%%%%%%%%%%%%%%%%%%%%%%%%%%%%%%%%%%%%%%%%%%%%%%

\end{document}